\documentclass[a4paper,twocolumn,11pt,accepted=2024-12-23]{quantumarticle}
\pdfoutput=1
\usepackage[utf8]{inputenc}
\usepackage[english]{babel}
\usepackage[T1]{fontenc}
\usepackage{amsmath}
\usepackage{mathtools}
\usepackage{braket}
\usepackage{hyperref}
\usepackage{stmaryrd}
\usepackage{subfigure}
\usepackage[shortlabels]{enumitem}
\usepackage[section]{placeins}
\usepackage{wasysym}
\usepackage[ruled,linesnumbered]{algorithm2e}
\usepackage{algpseudocode}
\usepackage{endnotes}

\begin{document}

\title{Minimising surface code failures using a color code decoder}

\author{Asmae Benhemou}
\affiliation{Department of Physics and Astronomy, University College London, London WC1E 6BT, United Kingdom}

\author{Kaavya Sahay}
\affiliation{Department of Applied Physics, Yale University, New Haven, Connecticut 06511, USA}

\author{Lingling Lao}
\affiliation{Department of Physics and Astronomy, University College London, London WC1E 6BT, United Kingdom}

\author{Benjamin J. Brown}
\affiliation{Niels Bohr International Academy, Niels Bohr Institute, Blegdamsvej 17, 2100 Copenhagen, Denmark}

\maketitle

\begin{abstract}
The development of practical, high-performance decoding algorithms reduces the resource cost of fault-tolerant quantum computing. Here we propose a decoder for the surface code that finds low-weight correction operators for errors produced by the depolarising noise model. The decoder is obtained by mapping the syndrome of the surface code onto that of the color code, thereby allowing us to adopt more sophisticated graph-based decoding protocols used for the color code. Analytical arguments and exhaustive testing show that the resulting decoder can find a least-weight correction for all weight $d/2$ depolarising errors for even code distance $d$. This improves the logical error rate by an exponential factor $O(2^{d/2})$ compared with decoders that treat bit-flip and dephasing errors separately. We demonstrate this improvement with analytical arguments and supporting numerical simulations at low error rates. Of independent interest, we also demonstrate an exponential improvement in logical error rate for our decoder used to correct independent and identically distributed bit-flip errors affecting the color code compared with more conventional color code decoding algorithms. 
\end{abstract}

\section{Introduction}

We envisage that a large-scale quantum computer will operate by performing fault-tolerant logic gates on qubits encoded using quantum error-correcting codes (QEC)~\cite{shor1995scheme, shor1996fault, calderbank1996good, terhal2015quantum, brown2016quantum, campbell2017roads}. As quantum information is processed, a classical decoding algorithm~\cite{dennis2002topological, duclos2010fast, wootton2012high, Anwar14, hutter2014efficient, Torlai2017neural, panteleev2021degenerate, Delfosse2021almostlineartime} will be used to interpret syndrome information that is collected by making parity measurements over the physical qubits of a code, to determine the errors it experiences. It is important to design high-performance and practical decoding algorithms to minimise the number of failure mechanisms that lead to a logical qubit error in real quantum systems. Such algorithms will reduce the high resource cost of encoding logical qubits, as better decoding algorithms will allow us to achieve a target rate of logical failures using fewer physical qubits. 

Topological codes, such as the surface code~\cite{kitaev2003fault, bravyi1998quantum, dennis2002topological, Fowler12surface} and color code~\cite{bombin2006topological, bombin2007topological, Bombin_2015_gauge, Kubica2015universal}, are among the most promising quantum error-correcting codes to be realised with quantum technology that is now under development~\cite{Egan2021fault, Ryan-Anderson2021realization, Krinner2022realizing, Sundaresan2023demonstrating, google2023suppressing, gupta2023encoding}. This is due to their layout which can be realised with a planar array of qubits with only nearest-neighbour interactions, and their demonstrated high threshold error rates, such that they can function under significant laboratory noise. A well-established class of decoders for the surface code~\cite{dennis2002topological, wang2003confinement, Raussendorf2007fault,  Fowler12surface} is based on the minimum-weight perfect-matching algorithm (MWPM)~\cite{edmonds1965paths, dennis2002topological, higgott2023sparse}. We colloquially refer to these decoders as matching decoders. Matching decoders give rise to high thresholds for topological codes, and have been demonstrated to be highly versatile in that they can be adapted to correct for different noise models~\cite{Fowler2013coping, Fowler2014quantifying,  Hutter2014breakdown, Nickerson2019analysingcorrelated, tuckett2020fault, bonilla2021xzzx, Strikis21, siegel2022adaptive, lin2023empirical}, as well as different codes~\cite{brown2020parallelized, nixon2021correcting, Miguel2023cellularautomaton}, including the color code~\cite{Wang2010graphical, delfosse2014decoding, Chamberland_2020, Beverland2021cost, sahay2022decoder, kubica2023efficient} and surface code variants~\cite{Raussendorf2005long, Higgott2021, Gidney2022benchmarkingplanar, Paetznick2023performance, kesselring2022anyon}. See the perspective article, Ref.~\cite{brown2022conservation} for a discussion.

A matching decoder exploits the structure of the code to identify errors which likely caused the error syndrome. Specifically, we take subsets of the syndrome data and represent them on a lattice, such that this syndrome data respects certain symmetries of the code~\cite{kitaev2003fault, brown2020parallelized, brown2022conservation}. These symmetries are characterised by the fact that errors produce syndrome violations, commonly known as defects, in pairs over the lattice. Given such a lattice, the MWPM algorithm can be used to pair spatially local defects. In turn, this pairing can be interpreted to identify an error that is likely to have given rise to the syndrome, such that a correction that recovers the encoded state can be determined.

The performance of a matching decoder is determined by the choice of syndrome lattice. The surface code has consistently demonstrated high thresholds by concentrating on syndrome lattices which correct for bit-flip errors and dephasing errors separately. However, without additional decoding steps~\cite{fowler2013optimal, delfosse2014decoding, higgott2023improved}, such decoders cannot identify correlations between these two error types that occur in common error models such as depolarising noise. See for example Refs.~\cite{duclos2010fast, fowler2013optimal} where candid discussions are presented on the limitations of matching decoder variations correcting for depolarising noise. In this work, we introduce a syndrome lattice for the surface code that accounts for correlations between bit-flip and dephasing errors, which we obtain by exploiting a correspondence between the surface code and the color code, commonly known as an unfolding map~\cite{bombin2012universal, Kubica_2015, Bhagoji15, criger2016noise}. Through this mapping, we can adopt color code decoding algorithms to correct noise on the surface code. Indeed, a syndrome lattice for our decoder naturally ensues by adapting the decoding methods introduced in Ref.~\cite{sahay2022decoder}.

We find that our unified decoder determines a least-weight correction for all weight $d/2$ depolarising noise errors for surface codes with even distance $d$. As such, in the limit of low error rate, we obtain optimal logical failure rates. In contrast, conventional matching decoders that do not account for correlated errors have a logical failure rate $O(2^{d/2})$ a factor higher than the decoder we present. We adopt a number of methods to analyse the performance of our decoder at low error rates. We first evaluate the number of weight $d/2$ errors that should lead to a logical failure for both our matching decoder, as well as a conventional matching decoder. We verify our expressions with exhaustive searching of weight $d/2$ errors. In addition, we measure the logical failure rate of our decoder using the splitting method~\cite{BENNETT1976, Bravyi_2013}. We find the numerical results we obtain to be consistent with our analytic results in the limit of vanishing physical error rate, thereby verifying the performance of our decoder in comparison to conventional decoders.
We also show that the threshold of our decoder is comparable to more conventional decoding methods. Finally, we investigate the performance of our scheme to decode the color code undergoing bit-flip noise. We find our analysis sheds light on the role of entropy in quantum error correction with the matching decoders that are central to our study~\cite{stace2010error, criger2018multi, beverland2019role}.

The remainder of this manuscript is structured as follows. In Sec.~\ref{sec:qec_codes}, we briefly review the quantum error-correcting codes of relevance in this study, namely the surface code and color code. In Sec.~\ref{sec:symmetries_and_decoding}, we review the symmetries of the color code, and how they are used to obtain matching decoders. We also explain why the unified decoder is capable of decoding high-weight errors. In Sec.~\ref{sec:mapping}, we describe the mapping from surface to color code, and introduce a noise mapping from depolarising errors on the surface code to a bit-flip noise on the color code. We present the results supporting the out-performance of the unified decoder over its restricted counterpart under depolarising noise on the surface code in Sec.~\ref{sec:results_toric_code}, and we present results using the same decoder to correct bit-flip noise on the color code in Sec.~\ref{sec:results_color_code}. We finally offer concluding remarks in Sec.~\ref{sec:discussion}.

\section{\label{sec:qec_codes} Preliminaries}

In this section we introduce the \textit{surface code} and the \textit{color code}, both of which are central to this work. We begin by stating the stabilizer formalism we use to describe these codes.

\subsection{The stabilizer formalism}

Quantum-error correcting codes are designed to protect logical information that is encoded in a subspace of a larger physical Hilbert space. This subspace is called the code space. A large class of QEC codes are readily described using the stabilizer formalism~\cite{Gottesman_1998}, whereby a QEC code is defined by an Abelian subgroup of the Pauli group $\mathcal{P}$, known as the stabilizer group $\mathcal{S} \subset \mathcal{P}$. 
Up to phases, the Pauli group on $n$ qubits is generated by the operators $n$-qubit Pauli matrices $X_q$ and $Z_q$ which respectively denote the standard Pauli-X and Pauli-Z operatorss acting on qubit $q$.
The code space of a stabilizer code is the +$1$ eigenspace of all of its elements, namely $s \ket{\psi} = \ket{\psi}$ $\forall$ $s \in \mathcal{S}$, where the code space is spanned by eigenstates $\ket{\psi}$. 

A stabilizer code is capable of detecting errors locally, without revealing information of the global encoded state. We consider Pauli errors $E$ that act on the code qubits. If an error $E$ anti-commutes with a subset of the stabilizer generators, when these are measured we obtain a set of syndrome defects to be used for error correction. We refer to a syndrome defect where a stabilizer measurement returns a negative outcome, i.e. $S_f E = - E S_f$.

A correction $C$ can be found and applied to return the code to the ground subspace such that $CE\in\mathcal{S}$, and recover the encoded logical state $\ket{\psi}$. We obtain $C$ using a decoding algorithm that is designed to determine a highly probable error giving rise to a measured syndrome, given a known error model where we assume errors are introduced at a low rate.  
In general, an $\llbracket n,k,d \rrbracket$ quantum code is characterised by the number of physical qubits $n$, the number of logical qubits $k$ that they encode, and a code distance $d$ given by the minimum weight of an error configuration triggering a logical error.

\subsection{\label{ssec:surface_code} The surface code}

\begin{figure}[t]
    \centering {\includegraphics[width=0.41\textwidth]{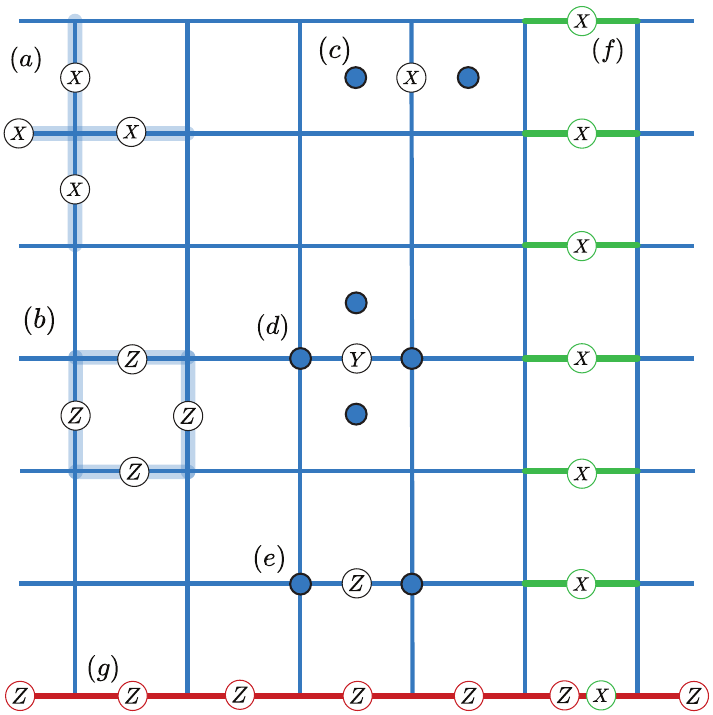}}
    \caption{The distance-$7$ unrotated surface code with qubits on the edges. (a) An $X$-type (star) parity check operator. (b) A $Z$-type (face) parity check operator. (c,d,e) Single-qubit $X$, $Y$ and $Z$ errors and their syndromes. (f) A vertical $\overline{X}$ logical operator. (g) A horizontal $\overline{Z}$ logical operator.}
    \label{fig:surface_code} 
\end{figure}

We define the surface code on a two-dimensional square lattice with boundaries, see Fig.~\ref{fig:surface_code}~\cite{kitaev2003fault, dennis2002topological}. A single physical qubit is supported on each edge $e$ of the lattice. The stabilizer group of the surface code is generated by vertex operators and face operators, denoted $A_v$ and $B_f$, respectively.
A vertex operator is defined at each site $v$ of the lattice, acting on the qubits located on the  incident edges $E_v$ on the vertex, such that $A_v = \Pi_{e\in E_v} X_e$. Similarly, we define an operator $B_f = \Pi_{e\in E_f} Z_e$ on each face $f$ of the lattice, where the set $E_f$ of edges is adjacent to the face $f$. 
On this planar variant of the surface code, the product of $Z$ ($X$) stabilizer operators creates a smooth (rough) boundary at the top and bottom (left and right) of the lattice. A string of Pauli operators terminating at the rough (smooth) boundaries generate logical $Z$ ($X$) operations on this encoded qubit, as shown in Fig.~\ref{fig:surface_code}., see Refs.~\cite{bravyi1998quantum, dennis2002topological} for details. 

In this work, we propose and implement a matching decoder to correct for the surface code undergoing an independent and identically distributed depolarising noise model. This error model is such that each qubit experiences either a Pauli-$X$, -$Y$, or -$Z$ error with probability $p/3$, or no error with probability $(1-p)$. We indicate examples of these single-qubit errors and their corresponding syndromes in Fig.~\ref{fig:surface_code}.

\subsection{\label{ssec:color_code} The color code}

\begin{figure}[t]
    \centering {\includegraphics[width=0.4\textwidth]{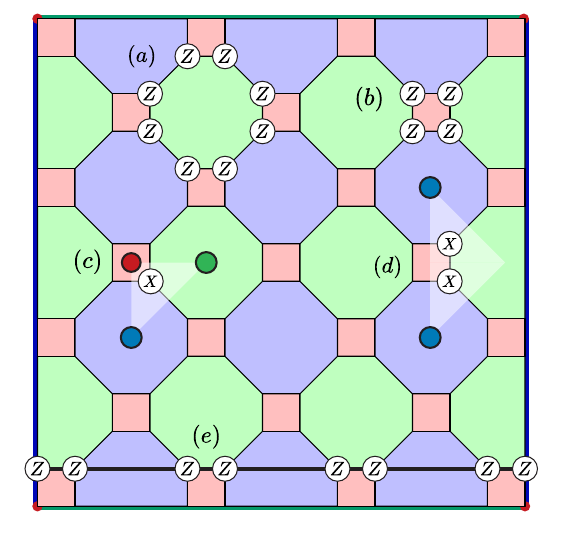}}
    \caption{The distance-$8$ square-octagonal color code with square boundaries. (a) An octagonal stabilizer generator. (b) A square stabilizer generator. (c) a single-qubit $X$-error. (d) A two-qubit $X$-error. (e) A horizontal logical operator.} 
    \label{fig:square_octagon_color_code} 
\end{figure}

We obtain a decoder for the surface code undergoing depolarising noise by mapping its syndrome onto that of the color code. Here we briefly describe the basic features of the color code, introduced in Ref.~\cite{bombin2006topological}.

The color code is defined on a three-colorable two-dimensional lattice. We will focus on the regular square-octagon color code with square boundaries, see Fig.~\ref{fig:square_octagon_color_code}.
Physical qubits are located on the vertices of the lattice, $v$, and each face $f$ supports two stabilizer operators $S_f^P = \prod_{v\in E_f} P_e$ where $P \in \{X, Z\}$ and $E_f$ denotes the qubits on the boundaries of face $f$. 

We define a set of color labels red, green and blue in bold $\mathcal{C} = \{\textbf{r}, \textbf{g}, \textbf{b}\}$, and each face of the code is assigned a label $\textbf{u} \in \mathcal{C}$ such that no two faces of the same color touch. 

We also assign colors to boundaries. The lattice of interest in Fig.~\ref{fig:square_octagon_color_code} has differently colored boundaries on its four sides and corners, outlined with their appropriate colors. To be precise, we attribute a color label to a boundary such that the qubits lying on a boundary of color $\textbf{u} \in \mathcal{C}$ do not support a face of color $ \textbf{u}$.  A corner on the lattice is attributed a color $\textbf{u}$ if its vertex only supports one face of color $\textbf{u}$. We find a $\textbf{u}$-colored corner at a vertex where two boundaries of colors $\textbf{v}$ and $\textbf{w}$ overlap where $\textbf{u} \neq \textbf{v} \neq \textbf{w} \neq \textbf{u}$. We note that a corner of color $\textbf{u}$ is also a member of its two adjacent boundaries that must necessarily have colors $\mathbf{v}$ and $\mathbf{w}$. 

The color code lattice we consider in Fig.~\ref{fig:square_octagon_color_code} features two green boundaries, two blue boundaries, and four red corners. We henceforth refer to the instance of the square-octagon color code shown in Fig.~\ref{fig:square_octagon_color_code} merely as the color code, and alternatively as the square-boundary color code when paying specific attention to its boundary conditions. 
 
The color code encodes two logical qubits with an even code distance $d$, using $n = 2(d-1)^2 + 2$ physical qubits. We define representatives of the logical operators of the code as follows:
\begin{equation}
    \overline{X}_{\textbf{u}} = \prod\limits_{v\in\delta\textbf{u}} X_v \text{   and   } \overline{Z}_{\textbf{u}} = \prod\limits_{v\in\delta\textbf{u}} Z_v
    \label{eq:logical_ops}
\end{equation}
where the product is taken over qubits lying on a boundary $\delta\textbf{u}$ of color $\textbf{u} = \textbf{g},\textbf{b}$, and obey the commutation relation $\overline{X}_{\textbf{u}}\overline{Z}_{\textbf{v}} = -\overline{Z}_{\textbf{v}}\overline{X}_{\textbf{u}}$ if and only if $\textbf{u} \not= \textbf{v}$. Otherwise, logical operators commute. These represent strings terminating at opposite boundaries of the same color, of which we show an example in Fig.~\ref{fig:square_octagon_color_code}(e). For now, we restrict our interest to the
Pauli-$Z$ stabilizer generators of the color code, and as such we omit the superscript label indicating the stabilizer type, such that $S_f = S_f^Z$. We show examples of an octagonal and a square stabilizer $S_f$ in Fig.~\ref{fig:square_octagon_color_code} (a) and (b) respectively. 
%We will consider a correspondence between depolarising noise on the surface and two-qubit bit-flip events on the color code. 
%These give rise to error strings $E=\prod_{v \in E_v}X_{v_1}X_{v_2}$, where $E_v$ is a subset of red stabilizers on the color code affected by two bit-flip errors on qubits $v_1$ and $v_2$.
We note that a defect is attributed the color from $\mathcal{C}$ of the violated stabilizer it lies on, and discuss the decoding of the color code in Sec.~\ref{sec:symmetries_and_decoding}.

\section{Materialised symmetries and decoding}\label{sec:symmetries_and_decoding}

\begin{figure}[t!]
    \centering
    \includegraphics[width=0.4\textwidth]{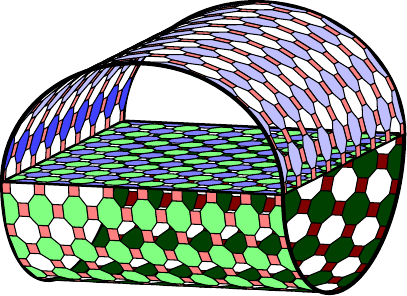}
    \caption{The manifold which preserves the global symmetry of the square-octagonal color code is a punctured torus.}
    \label{fig:488_symmetry}
\end{figure}

In this section, we review the concept of materialised symmetries for topological quantum codes introduced in Refs.~\cite{kitaev2003fault, brown2020parallelized, sahay2022decoder,brown2022conservation}. These symmetries give rise to a parity conservation law on syndrome defects, that is, any error creates an even number of defects on the symmetry. We exploit this parity conservation law to make a matching decoder for the square-boundary color code. In Ref.~\cite{sahay2022decoder}, a global symmetry termed the \textit{unified lattice} was derived for the triangular color code with a single red, blue and green boundary. Notably, the unified lattice presented in Ref.~\cite{sahay2022decoder} had the topology of a M\"obius strip. Due to the unified lattice's improved capacity to handle correctable errors at the boundary~\cite{gidney2023newcircuitsopensource},  Ref.~\cite{sahay2022decoder} reported improvements in both the threshold and logical failure rates for the triangular color code using decoding on the unified lattice. Motivated by this, we introduce a unified decoder for the color code presented in Sec.~\ref{ssec:color_code}. In addition to introducing the unified lattice decoder, we will also review the matching decoder on the restricted lattice~\cite{Wang2010graphical, delfosse2014decoding, Chamberland_2020, Beverland2021cost, kubica2023efficient}. In later sections, we will compare the restricted-lattice decoder to the unified-lattice decoder.

\subsection{\label{ssec:symmetries} Symmetries of the color code}

We use symmetries to obtain matching decoders. We define a symmetry as a subset of stabilizers $\Sigma$ such that  
 \begin{equation}
   \prod_{s \in \Sigma} s = \mathbf{1}.
 \end{equation}
This definition is such that the elements of the symmetry $\Sigma$ necessarily give rise to an even number of syndrome defects under any error. Given an appropriate choice of $ \Sigma $ we can thus design a matching decoder, demonstrated with a number of examples~\cite{dennis2002topological, Wang2010graphical, delfosse2014decoding, Nickerson2019analysingcorrelated,  tuckett2020fault, bonilla2021xzzx, nixon2021correcting,Srivastava2022xyzhexagonal, sahay2022decoder, kubica2023efficient, Miguel2023cellularautomaton,  huang2022tailoring} which are summarised in Ref.~\cite{brown2022conservation}.

We find the symmetries of the color code by considering subsets of its stabilizer generators with specific colors \cite{delfosse2014decoding,sahay2022decoder}. %These subsets are known as restricted lattices, as we will define below.
Our goal is to derive a single `unifying' symmetry for the color code with boundaries. We also investigate other color code decoders based on restricted lattices. Interestingly, we find that the square-boundary color code has a unified lattice with a topology that is distinct from that in Ref.~\cite{sahay2022decoder}, and is given by the manifold in Fig.~\ref{fig:488_symmetry}. In this subsection, we show how to derive this unified lattice for the square boundary color code, which we will adopt in our matching decoder implementation.

\begin{figure*}[ht]
    \centering
    \includegraphics[width=0.98\textwidth]{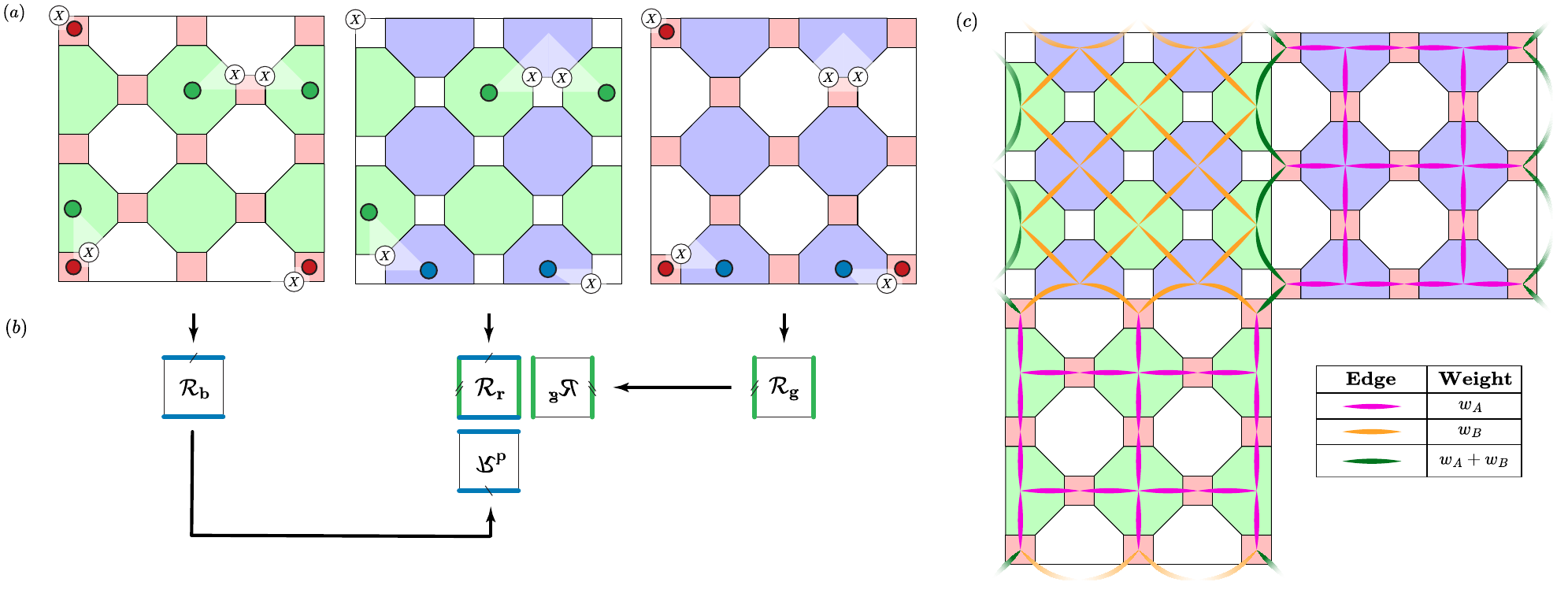} 
    \caption{(a) One and two-qubit errors with the corresponding syndromes on the three restricted lattices of the $d=6$ color code. (b) Stitching the lattices together along the boundary operators. (c) Edge weights for the matching graph. This configuration is empirically chosen to correct for all expected minimum-weight uncorrectable errors, but is non-unique.}
    \label{fig:restricted_color_code}
\end{figure*}

Prior work has established a specific set of bulk symmetries, referred to as restricted lattices on the color code~\cite{delfosse2014decoding}. 
A $\mathbf{uv}$-restricted lattice, denoted by $\mathcal{R}_\mathbf{w}$ (where $\textbf{u} \neq \textbf{v} \neq \textbf{w} \neq \textbf{u}$), is defined as
\begin{equation}
    \mathcal{R}_{\mathbf{w}} = \{ \{S_f \} _{\mathbf{col}(f) \neq \mathbf{w} } \}.
\end{equation}
We show the restricted lattices $\mathcal{R}_\mathbf{b}, \mathcal{R}_\mathbf{r},$ and $ \mathcal{R}_\mathbf{g}$ in Fig.~\ref{fig:restricted_color_code}(a). Errors in the bulk of the color code preserve defect parity on any single restricted lattice. We show examples of bulk errors on the restricted lattices in Fig.~\ref{fig:restricted_color_code}(a), where the bottom left error on the lattice flips a single qubit and creates a single defect of each color, and the top right error flips two qubits, and creates two green defects.

Note that the restricted lattices are not symmetries for the color code with boundaries. Errors at the boundary may give rise to a single defect on the restricted lattice, see the error in Fig.~\ref{fig:restricted_color_code}(a) (bottom right). We must therefore introduce a boundary operator to complete the restricted lattice symmetry; see Refs.~\cite{sahay2022decoder, brown2022conservation} for discussions on boundary operators.

We introduce a boundary operator for each restricted lattice, defined as
\begin{equation}
    b_{\mathbf{u}} = \prod_{\mathbf{col}(f) \neq \mathbf{w} } S_f
\end{equation}
where we take the product over all faces of the restricted lattice, i.e. those not colored $\mathbf{u}$. This operator is highlighted for each restricted lattice in Fig.~\ref{fig:restricted_color_code}(b). We note that the boundary operator $b_{\mathbf{u}}$ is supported on all boundaries and corners that are not $\mathbf{u}$-colored.

We find that the inclusion of the boundary operator with its restricted lattice gives rise to a materialised symmetry~\cite{sahay2022decoder}. 
A set that includes this boundary operator, $b_\mathbf{u}$, together with the face operators of its respective restricted lattice, $\mathcal{R}_\mathbf{u}$, completes a symmetry and thereby allows us to match lone defects on the restricted lattice to the boundary. Explicitly, we write this symmetry as follows
\begin{equation}
   \Sigma_{\mathbf{u}} = \{b_{\mathbf{u}} , \mathcal{R}_{\mathbf{u}} \}  , \quad \prod_{s \in \Sigma_{\mathbf{u}}} s = \mathbf{1}.
\end{equation}

We now derive a global symmetry by taking products of smaller symmetries. Relationships among the boundary operators reveal that restricted lattices can be unified by their boundaries, $b_{\mathbf{r}}b_{\mathbf{g}}b_{\mathbf{b}} = {\mathbf{1}}$. This is discussed in generality in Ref.~\cite{sahay2022decoder}. Let us describe this unification procedure in steps. We start by examining the subproduct $b_{\mathbf{r}} b_{\mathbf{g}}$ of this equation; this is essentially the combination of the $\mathbf{gb}$ and $\mathbf{rb}$ restricted lattices, as in Fig.~\ref{fig:restricted_color_code}(b) (centre right). Because the qubits at the right (left) of $\mathcal{R}_{\mathbf{r}}$ are the same as those on the left (right) of a horizontally flipped $\mathcal{R}_{\mathbf{g}}$, this product stitches the symmetries together, thus forming a cylinder. We now add $b_{\mathbf{b}}$, the product of all faces on $\mathcal{R}_b$ to this subproduct. Analogously, the qubits at the top (bottom) of $\mathcal{R}_{\mathbf{r}}$ are the same as that on the bottom (top) of a vertically flipped $\mathcal{R}_{\mathbf{b}}$, and we join them  accordingly, as in Fig.~\ref{fig:restricted_color_code}(b) (left). This creates the manifold seen in Fig.~\ref{fig:488_symmetry}.

We now provide an alternate explanation of the stitching procedure adopted to produce the unified lattice. To do so, we consider the syndromes created by errors at the lattice boundary. Let us specifically consider a single qubit error at the lattice boundary, such as the single error shown at the green boundary in Fig.~\ref{fig:restricted_color_code}(a) (bottom right). We concentrate our attention on the two single defects created by this error on both the $\mathcal{R}_\mathbf{r}$ and $\mathcal{R}_\mathbf{b}$ lattices. Ideally, these defects should be paired by a single low-weight edge. As such, we can consider stitching these two restricted lattices together along the green boundary, such that these two defects are adjacent on the unified lattice. Next, by analogy, we join $\mathcal{R}_\mathbf{r}$ with $\mathcal{R}_\mathbf{g}$ along the $\mathbf{b}$ boundaries. The system takes on the topology of a torus with a disk-shaped puncture, again resulting in the manifold of Fig.~\ref{fig:488_symmetry}. 

Let us additionally examine a single qubit error at the corner of the lattice to show how we stitch the corners of the unified lattice. See for example the top left error in Fig.~\ref{fig:restricted_color_code}(a). This error produces syndromes only on $\mathcal{R}_\mathbf{b}$ and $\mathcal{R}_\mathbf{g}$, not on $\mathcal{R}_\mathbf{r}$. Consequently, we would like our matching decoder to pair defects across any corner of $\mathcal{R}_\mathbf{b}$ to the equivalent corner of  $\mathcal{R}_\mathbf{g}$. This manifests on the unified lattice as a string stretching from the top of the cylinder diagonally across to the sheet-like handle.

\subsection{\label{ssec:decoding_on_unified_lattice} Matching decoders and code symmetries}

We can use symmetries to design decoders based on minimum-weight perfect matching. See Ref.~\cite{brown2022conservation} for a discussion. Here, let us briefly review the restricted-lattice decoder and the unified lattice decoder that we use throughout this work.

\begin{figure}[ht!]
    \centering{
    \includegraphics[width=0.48\textwidth]{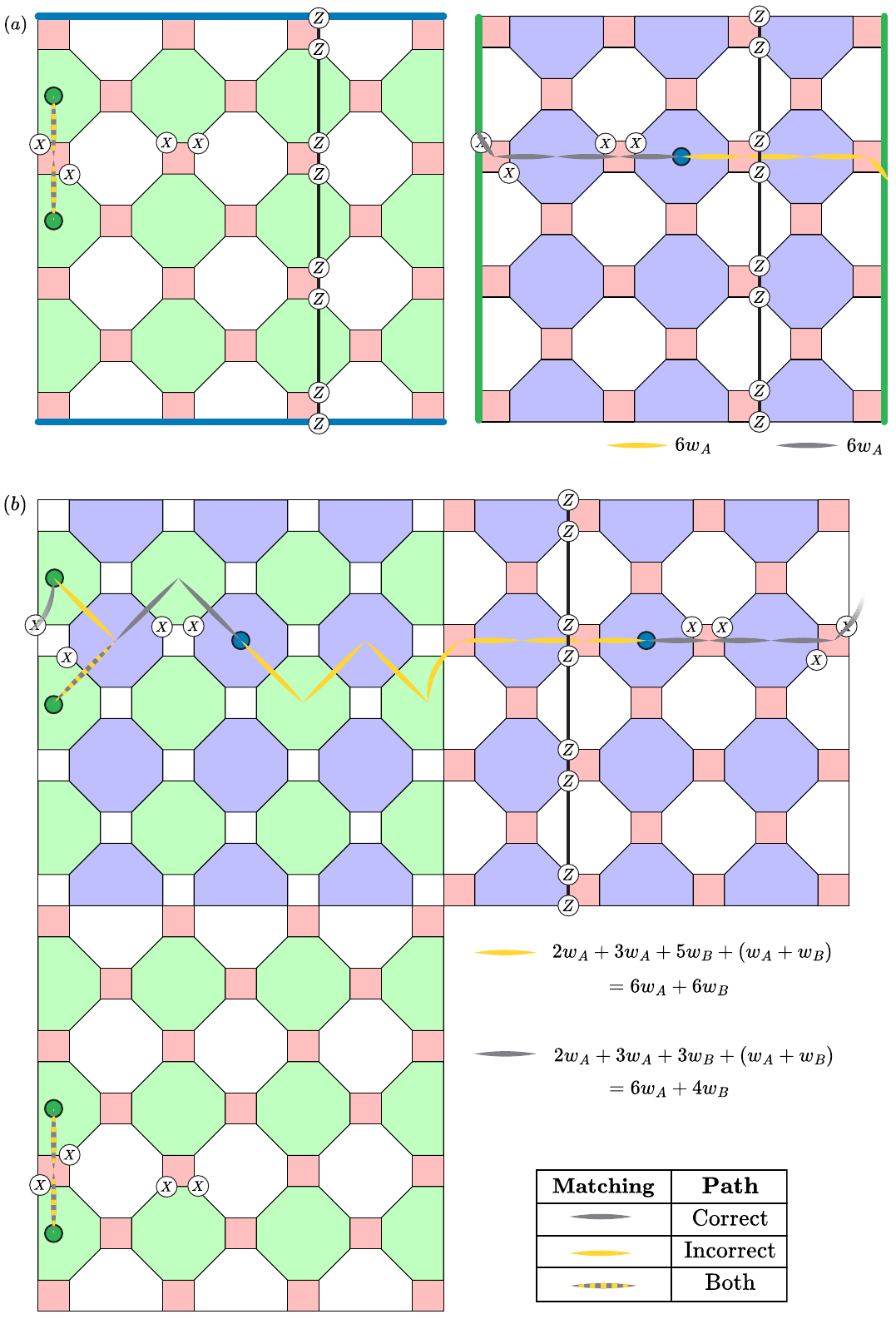}}
    \caption{Here we consider a weight-$4$ error on a $d = 8$ color code lattice, which can be interpreted as a weight-$2$ error on the $d = 4$ surface code as per Sec.~\ref{sec:mapping}, further shown in Fig.~\ref{fig:toric_cc_mapped}. We examine correction operators found by matching using (a) the restricted decoder, and (b) the unified decoder.  A correct (incorrect) pairing of defects that returns the encoded qubit to the correct (incorrect) state in the logical subspace is depicted in grey (yellow). Matching on the restricted lattice is unable to reliably determine the correct defect pairing. Matching on the unified lattice correctly pairs the syndrome defects when the error string contains a Pauli-$Y$ error, by having a demonstrably lower weight.}
    \label{fig:example_of_unified_decoder_success}
\end{figure}

Prior work has proposed using the individual restricted lattices for error correction \cite{kubica2023efficient}.
In the restriction-lattice decoder, defects on each of the $\mathcal{R}_{\mathbf{g}}$ and $\mathcal{R}_{\mathbf{b}}$ lattices are matched within the lattice or to the boundary operator that completes the symmetry to produce local correction operators $\{ \mathcal{C}_{\mathbf{g}} , \mathcal{C}_{\mathbf{b}} \}$. In Fig.\ref{fig:example_of_unified_decoder_success}(a), for instance, these consist of Pauli corrections on the qubits lying along the highlighted paths. In order to apply a global correction, the union of the local correction operators $\mathcal{C}_{\mathbf{g}} \cup \mathcal{C}_{\mathbf{b}}$ is applied to the underlying physical qubits.

The unified lattice is obtained from the global symmetry of the color code described in the previous subsection, on which errors give rise to pairs of defects. 
Given a collection of defects on the unified lattice, the matching-based decoding problem reduces to finding a pairing between these defects that is created by an underlying error with maximal probability. 
Towards this goal, a matching between a pair of defects is assigned a weight inversely dependent on the probability of errors on all the qubits along the shortest path connecting the two. 
As a result, decoding is equivalent to finding the perfect matching of minimum weight on this weighted graph on the unified lattice.

It was seen in  Ref.~\cite{sahay2022decoder} that the choice of edge weights on this graph can be determined by examining simple error patterns at the bulk, boundary and corners of the unified lattice. We apply the same basic concept in this work, but we set the edge weights as free parameters $w_A$ to bulk edges on $\mathcal{R}_\mathbf{g}$ and $\mathcal{R}_\mathbf{b}$, and $w_B$ on $\mathcal{R}_\mathbf{r}$, as shown in Fig~\ref{fig:restricted_color_code}(c), where we adopt a matching graph which allows for the correction of all expected errors when the global symmetry is respected. In the following sections, specific weights were found by numerical tests depending on the noise model.   
 
Given a matching on a particular lattice -- unified or restricted -- we now lay out the procedure to obtain the relevant correction to the logical qubit subjected to $X$ errors on the underlying physical qubits. Note that the following procedure is valid for both the restricted and unified lattices. We first apply the Pauli-$Z$ corrections to qubits along the shortest paths determined by the matching algorithm. This returns the logical qubit to the code space. During this process, we track the parity of a logical operator, here the vertical $Z$ logical depicted in Fig.~\ref{fig:example_of_unified_decoder_success}. We successfully return the encoded qubit to the correct logical state if the decoding algorithm has determined that the parity of matched paths \textit{across} this logical is the same as the parity of physical Pauli errors on the qubits \textit{along} the logical, a method discussed in more detail in Ref.~\cite{brown2022conservation, sahay2022decoder}. For example, in Fig.~\ref{fig:example_of_unified_decoder_success}, if the decoder finds the yellow matching, the corresponding correction, together with the initial error, applies $\overline{X}$ to the encoded qubit; the grey matching returns the lattice to the correct state. Henceforth, we will refer to the MWPM-based decoder applied to syndrome data on the restricted (unified) lattice as the restricted (unified) decoder.

% \section{\label{sec:mapping} Mapping from surface to color code}
\section{\label{sec:mapping} Mapping surface code errors to color code errors}

\begin{figure*}[ht]
    \centering
    {\includegraphics[width=0.99\textwidth]{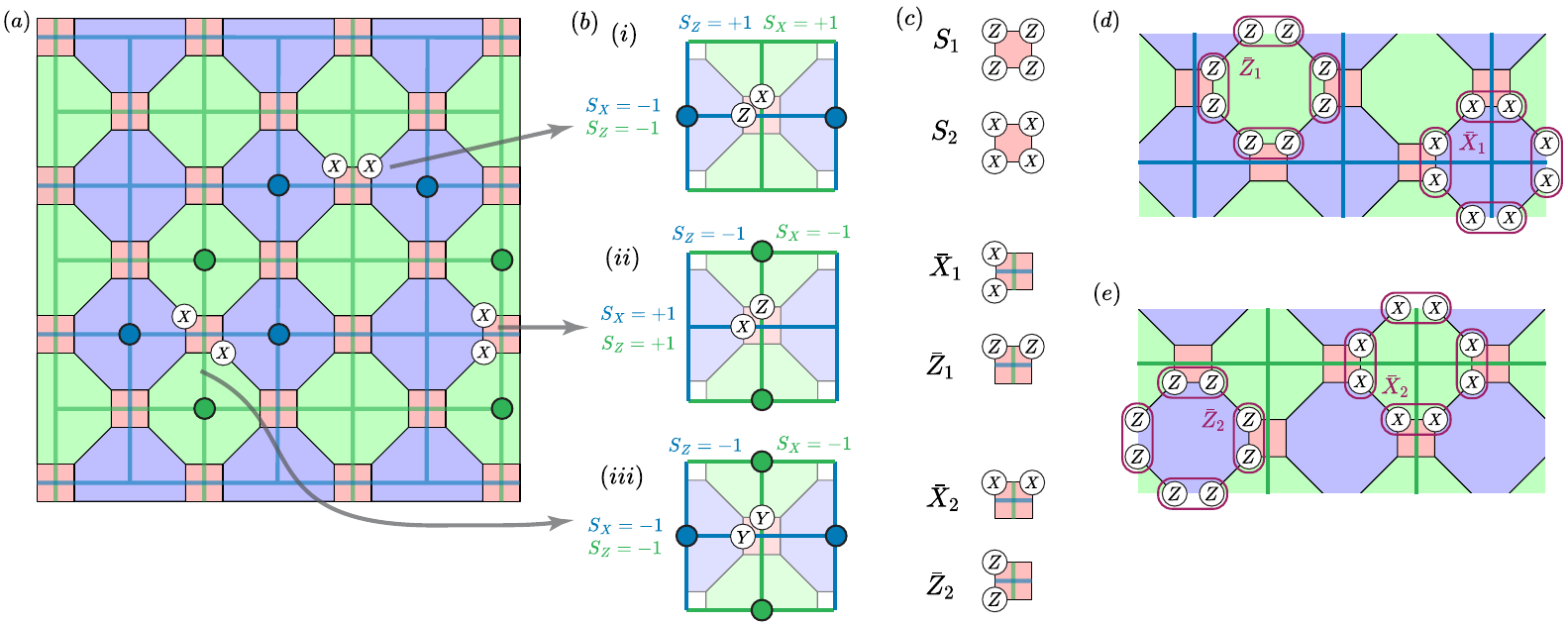}}
    % {\includegraphics[width=0.9\textwidth]{figures/new_schematics/SOCC_mappingCombined.pdf}}
    \caption{The surface-color code mapping. (a) The color code with an overlay of two surface codes - blue and green - to indicate their positioning with respect to the color code stabilizers, along with weight-$2$ bit-flip errors and their syndromes on the color code. (b) Correspondence of the indicated weight-$2$ bit-flip errors on the color code to single-qubit depolarising noise on the surface codes. For a given error on the color code in (a), the corresponding error for the blue (green) surface code is shown on the blue (green) edge across the red squares in (i-iii), with the surface code stabilizer measurements to its left and top displayed in the same color. (c) Stabilizers and logical operators of the $\llbracket 4,2,2\rrbracket$ code defined on a single square with qubits on its vertices. The displayed orientation is used for a red square traversed by a horizontal (vertical) edge of the blue (green) surface code. In the alternate case, an orientation rotated by $90^{\circ}$ applies. (d,e) Stabilizers of the (d) blue and (e) green surface codes obtained from the mapping in Sec.~\ref{sec:mapping}. The star operators of the blue (green) surface code are Pauli-$X$ stabilizers, where each $X$ operator is encoded as the logical $\overline{X}_1(\overline{X}_2)$ operator of the $\llbracket 4,2,2\rrbracket$ codes shown in (c). The face operators are Pauli-$Z$ stabilizers where each $Z$ operator is encoded as the logical $\overline{Z}_1(\overline{Z}_2)$ operator of the $\llbracket 4,2,2\rrbracket$ codes.}
\label{fig:toric_depolarising_error_mapping}
\end{figure*}

We have introduced the surface code and the color code in the previous section, and proposed a decoding procedure for bit-flip errors acting on the color code by matching on either the restricted lattice or the unified lattice. We now describe the local mapping from the surface code onto the square-octagonal color code, inspired by the unfolding of the color code \cite{Kubica_2015}, and introduce a mapping from depolarising errors on the former to correlated bit-flip errors on the latter. We exploit this mapping to adopt the restricted-lattice and unified-lattice color code decoder for the surface code undergoing depolarising noise.

% We now describe a local mapping from the surface code onto the square-octagonal color code and introduce a mapping from depolarising errors on the former to correlated bit-flip errors on the latter. We exploit this mapping to adopt the restricted-lattice and unified-lattice color-code decoder for the surface code undergoing depolarising noise. 

The two-dimensional color code is locally equivalent to two copies of the surface code~\cite{yoshida2011classification, bombin2012universal, Bhagoji15, Kubica_2015, criger2016noise}. We first describe this equivalence via operators, before providing a formalism to reconstruct it. As a guide for the mapping, we show the surface code in blue overlaid with the color code in Fig.~\ref{fig:toric_depolarising_error_mapping}(a). The figure additionally depicts a second copy of the surface code in green, laid out on the dual lattice, that duplicates the syndrome of the blue copy. Considering two copies of the duplicate syndrome facilitates the unfolding map between two copies of the surface code and the color code.

Let us begin by stating the equivalence between the different lattice elements under this mapping, by considering the surface code displayed with a blue lattice in Fig.~\ref{fig:toric_depolarising_error_mapping}(a). First, we map the vertex and face operators of this blue surface code onto octagonal face operators of the color code. Specifically, each surface code star operator corresponds to a blue octagon $S_f$ of the color code and, likewise, each face operator of the surface code corresponds to a green octagonal face operator in the color code picture $S_f$. This is shown in Fig.~\ref{fig:toric_depolarising_error_mapping}(a), where the star and face operators of the blue surface code align with their corresponding color code stabilizers exactly, thereby illustrating this mapping. We also show the surface code displayed with a green lattice, which represents a rotation of the blue counterpart, and where a similar argument holds.

Now that we have identified stars and plaquettes of each surface code with octagonal face operators of the color code, we must choose errors in the color code picture that give rise to equivalent syndromes in the surface code picture. A Pauli-$Z$ error on the surface code will violate its two adjacent vertex operators, as shown in Fig.~\ref{fig:surface_code}. In the color code picture, this error will violate the corresponding blue octagonal operators as shown in  Fig.~\ref{fig:toric_depolarising_error_mapping}(b)(i) where we indicate this $Z$-flip on the blue surface code edge. To obtain this violation in the color code picture, we add two horizontal bit-flip errors onto the single red face of the color code on which the edge lies such that its two adjacent blue octagons support a syndrome, see Fig.~\ref{fig:toric_depolarising_error_mapping}(a). Similarly, a Pauli-$X$ error in the surface code picture will violate two adjacent face operators, which correspond to green octagons in the color code picture in Fig.~\ref{fig:toric_depolarising_error_mapping}(b)(ii). We find that the error in the color code picture that violates these appropriate color code octagons is a pair of bit-flip errors lying vertically on a single red square-shaped face operator shown in the corresponding diagram in Fig.~\ref{fig:toric_depolarising_error_mapping}(a). 

We can obtain a Pauli-$Y$ error by adding a Pauli-$Z$ error and a Pauli-$X$ error to the same red face of the color code, which results in a diagonal bit-flip pair on the latter. Since the surface code defined using the green lattice is rotated, so is the orientation of the errors it detects, namely a horizontal(vertical) pair of bit-flips on such a red plaquette in the color code picture corresponds to a Pauli-$X(Z)$ error, also shown in Fig.~\ref{fig:toric_depolarising_error_mapping}(a) and (b). 

We have now proposed an error model that violates pairs of either blue or green face operators according to the mapping from the surface code undergoing a depolarising noise model. As we have alluded, single qubit errors in the surface code picture correspond to two-qubit errors supported on individual red faces of the color code. Indeed, we see that the edges in the surface code picture each have an underlying red face operator in the color code picture, where the red square inherits an orientation from the underlying edge. We therefore identify qubits in the surface code picture with red faces in the color code picture. In what follows, it will be helpful to keep this identification in mind as we describe the mapping more rigorously. We note that this mapping is non-trivial in the sense that the commutation relations of the errors are not preserved. Nevertheless, we find a valid decoder for the surface code with the equivalence we have proposed.

Let us make this mapping more concrete using the more rigorous language of the unfolding map~\cite{bombin2012universal,Bhagoji15, Kubica_2015}. Specifically, we duplicate the surface code supporting the syndrome, where the second copy differs from the first by a transversal Hadamard rotation. We show the duplicate surface code syndrome on the dual lattice in green in Fig.~\ref{fig:toric_depolarising_error_mapping}.
We then put the two duplicate copies through a folding map, see e.g. Refs.~\cite{bombin2012universal, Kubica_2015, criger2016noise}. Specifically, we will consider the unfolding map of Ref.~\cite{criger2016noise} where pairs of surface code qubits are encoded with the $\llbracket4,2,2 \rrbracket$-code. See also Ref.~\cite{Brell_2011} on this equivalence. This four-qubit code can be regarded as an inner code for a concatenated model with the surface code as the outer code. All together, this concatenated model gives the square-octagon color code we have studied throughout this work.

Under this mapping, the inner $\llbracket4,2,2 \rrbracket$ code can be regarded as a red face of the color code. Indeed each disjoint red face supports the four qubits of the $\llbracket4,2,2 \rrbracket$ code lying on its vertices as shown in Fig.~\ref{fig:toric_depolarising_error_mapping}(c). Furthermore, the stabilizer group of the inner code is generated by $S^X = XXXX$ and $S^Z = ZZZZ$, which coincide with the red face operators of the square-octagon color code. Moreover, the inner code encodes two logical qubits represented by the operators $\overline{X}_1 = XIXI$, $\overline{Z}_1 = ZZII$, $\overline{X}_2 = XXII$ and $\overline{Z}_2 = ZIZI$ visualised per the convention in Fig.~\ref{fig:toric_depolarising_error_mapping}(c).

We then express the stabilizers of the blue and green surface codes respectively, using the two logical operators of the inner code $\{\overline{X}_1, \overline{Z}_1\}$ and $\{\overline{X}_2, \overline{Z}_2\}$, where the first(second) qubit encodes the blue(green) copy of the surface code, as shown in Fig.~\ref{fig:toric_depolarising_error_mapping}(d) (Fig.~\ref{fig:toric_depolarising_error_mapping}(e)). We emphasise that that the orientation of the edge qubit of the surface code dictates the orientation of the encoding of the $\llbracket4,2,2 \rrbracket$ code. See Fig.~\ref{fig:toric_depolarising_error_mapping}(c). Indeed, a red square associated with a horizontal (vertical) edge of the blue (green) surface code is encoded as per Fig.~\ref{fig:toric_depolarising_error_mapping}(c), but when traversed by a vertical edge, this encoding must be rotated by 90\textdegree~to preserve our mapping. 

Hence, in this mapping, a Pauli operator on a surface code qubit corresponds to a pair of Pauli operators on the red stabilizer lying on the associated edge, where this pair defines the appropriate logical operator of the $\llbracket 4,2,2 \rrbracket$ code. This is summarised in the following equivalences between stabilizers of the surface codes, color codes, and product of logical operators of the inner code 
\begin{align}        
    S_{\text{SC,}\mathbf{b}}^X &\equiv S_{f, \mathbf{b}}^{X} \equiv \prod_{r\in\partial{f}}\overline{X}_1^r \\
    S_{\text{SC,}\mathbf{b}}^Z &\equiv S_{f, \mathbf{g}}^{Z} \equiv \prod_{r\in\partial{f}}\overline{Z}_1^r \\
    S_{\text{SC,}\mathbf{g}}^X &\equiv S_{f, \mathbf{g}}^{X} \equiv \prod_{r\in\partial{f}}\overline{X}_2^r \\
    S_{\text{SC,}\mathbf{g}}^Z &\equiv S_{f, \mathbf{b}}^{Z} \equiv \prod_{r\in\partial{f}}\overline{Z}_2^r 
    \label{eq:stabilizer_relations}
\end{align}
where the product over $r\in\partial{f}$ indicates that the logical operator is applied to all four red square operators on the boundary of the appropriate color code stabilizer. 

Finally, since the surface code on the green lattice is a duplicate of its blue counterpart up to a Hadamard rotation, we identify certain logical operators of the inner code to complete our mapping. Indeed, under the duplication, with the Hadamard rotation, we have the following relations between pairs of logical operators of the inner code, namely $\overline{X}_1 \equiv \overline{Z}_2$ and $\overline{Z}_1 \equiv \overline{X}_2$. This ensures the syndrome correspondence in Fig.~\ref{fig:toric_depolarising_error_mapping}(a) such that all depolarising errors on the surface code now have a one-to-one correspondence with a commuting two-qubit bit-flip error model. Most importantly, this enables us to use a decoder for a CSS code \cite{calderbank1996good, steane1996simple} to correct after depolarising noise on the surface code, represented in the color code picture.

\section{\label{sec:results_toric_code} Surface code performance under depolarising noise}

\begin{figure}[t]
    \centering
    \includegraphics[width=0.17\textwidth]{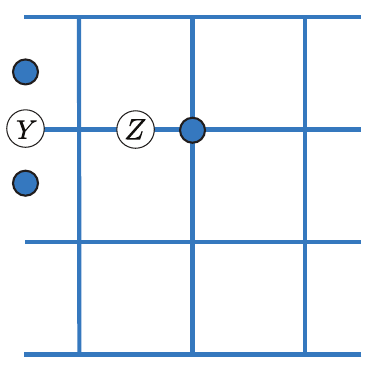}
    
    \caption{The $d=4$ surface code, with depolarising errors giving rise to the corresponding set of syndrome defects seen in Fig.~\ref{fig:example_of_unified_decoder_success} on the color code.}
\label{fig:toric_cc_mapped}
\end{figure}

In this section, we study the surface code under depolarising Pauli noise on data qubits and perfect measurements. We compare the performance of the surface code decoder we have introduced to the conventional matching decoder. Under the mapping, these correspond to the unified decoder and the restricted decoder, respectively.

Our analysis begins by counting the number of minimum-weight uncorrectable errors on the lattice for each decoder. The unified decoder provides an advantage that results in a logical error rate suppression in the low physical error-rate regime. We interrogate our analytical estimates numerically using the splitting method~\cite{Bravyi_2013}. Additionally, we investigate the threshold performance of the code numerically using \texttt{PyMatching}~\cite{higgott2023sparse}.

\subsection{\label{ssec:motivation} Correcting Pauli-Y errors on the surface code}

Under depolarising noise, $p(Y)=p(X)=p(Z)$. Since $Y=XZ$, breaking $Y$s down  into $X$ and $Z$ creates correlated $X$ and $Z$ errors and corresponding correlated syndromes among the star and plaquette stabilizers. However, these correlations generally elude decoders which treat $X$ and $Z$ errors independently. As we will argue, the unified decoder identifies these correlations.

We use the unified decoder on the syndrome of the surface code mapped onto the color code lattice to correct for a depolarising noise model. This is motivated by the ability of such a decoder to identify Pauli-$Y$ errors on the surface code, since each one gives rise to a distinct syndrome pattern on the unified lattice. Indeed, a Pauli-$X$ error on the blue surface code introduces a syndrome defect on $\mathcal{R}_{\mathbf{r}}$ and $\mathcal{R}_{\mathbf{b}}$ as shown in Fig.~\ref{fig:toric_depolarising_error_mapping}. Similarly, a Pauli-$Z$ error leads to a syndrome on $\mathcal{R}_{\mathbf{r}}$ and $\mathcal{R}_{\mathbf{g}}$, and finally a Pauli-$Y$ error is distinguished by its syndrome featuring on all three restricted lattices. Hence, we find that in the presence of at least one $Y$ error, decoding on the unified lattice captures correlations that any one or pair of restricted lattices remain oblivious to.

To illustrate our argument, we provide an example of a minimum-weight uncorrectable error on the surface code with one Pauli-$Y$ that corresponds to a weight $d/2$ bit-flip error on the color code in Fig.~\ref{fig:example_of_unified_decoder_success}. Note that due to the presence of a diagonal bit-flip pair on a red stabilizer of the color code, the syndrome on the unified decoder spans all three sub-lattices and a lower-weight matching path emerges for the successful correction (Fig.~\ref{fig:example_of_unified_decoder_success}(a)). 
We numerically find that this improvement from the unified decoder is valid for a range of values of weights $w_A$ and $w_B$ in the matching graph depicted in Fig.~\ref{fig:restricted_color_code}(c). For example, $w_A = w_B = 1$ achieves an optimal number of successful corrections.

\subsection{\label{ssec:path_counting_toric} Path-counting of failure mechanisms}

The advantages of using the unified-lattice decoder are very apparent in the limit where physical error rates $p$ are low. In the regime of asymptotically low $p$, the logical failure rate is dominated by minimum-weight failure mechanisms, such that the following holds
\begin{equation}
    \overline{P}_{0,n} := \underset{p \rightarrow 0}{\text{lim}}\overline{P}(p,n) \sim N_{\textrm{fail}} p^{d/2}
\end{equation}
where $N_\textrm{fail}$ is the entropic term that denotes the number of least-weight errors leading to a logical failure.

In what follows we evaluate the entropic term for both the restricted-lattice decoder and the unified lattice decoder, $N^{\textrm{res}}_{\textrm{fail}}  $ and $ N^{\textrm{uni}}_{\textrm{fail}} $ respectively. We find the restricted-lattice decoder has an exponentially larger number of failure mechanisms in the low error rate regime than the unified-lattice decoder as a function of the code distance:
\begin{equation}
 N^{\textrm{res}}_{\textrm{fail}} / N^{\textrm{uni}}_{\textrm{fail}} = 2^{d/2}. \label{eqn:ratio}
\end{equation}
 We attribute this factor to the ability of the unified-lattice decoder to identify the occurrence of Pauli-Y errors. First, we count $N^{\textrm{res}}_{\textrm{fail}}  $ and $ N^{\textrm{uni}}_{\textrm{fail}} $ exactly. In the following section we interrogate the relative performance of the two decoders numerically, and verify our expressions with exhaustive testing.

We now evaluate the number of least-weight errors that lead the surface code to failure using the restricted-lattice decoder. This decoder cannot distinguish between Pauli-Z or Pauli-Y errors. As such we count the number of weight $d/2$ configurations of errors along a major row of the color code lattice, where we say a major row is a row of $d$ red plaquettes. In contrast, a minor row of the lattice is a row of $d-1$ red color code plaquettes. Since the lattice is square, it supports $d$ major rows and $d-1$ minor rows. 

The probability of $d/2$ Pauli-Z or Pauli-Y errors occurring is $(1-p)^{n - d/2} (2p/3)^{d/2}$. There are $d$ major rows on which one of these error configurations can occur, and on each of these rows, there are $d! \times (d/2 )!^{-2} $ distinct configurations the $d/2$ errors can find on the $d$ plaquettes of the major row. We therefore find that 
\begin{align}
     \overline{P}^{\textrm{res}}_{0,n} &= (1-p)^n  \frac{d}{2}  \binom{d}{\frac{d}{2}} \left(\frac{2p }{3 (1-p)}\right)^{\frac{d}{2}} \\ &\approx  \frac{d}{2}  \binom{d}{\frac{d}{2}} \left(\frac{2}{3}\right)^{ \frac{d}{2}} p ^{\frac{d}{2}}  .
\end{align}
where we include an additional prefactor of one half due to the matching decoder correctly guessing a correction that successfully corrected the error with probability $1/2$ and, on the right hand side, we assume that $1-p \approx 1$ in the limit of very small $p$. Expressing the equation this way allows us to read off the entropic term
\begin{equation}
    N^{\textrm{res}}_{\textrm{fail}} = \frac{d}{2} \binom{d}{\frac{d}{2}} (2/3)^{d/2}.
    \label{eq:N_fail_toric_restricted_even}
\end{equation}

 We verify Eq.~\ref{eq:N_fail_toric_restricted_even} by exhaustively testing the restricted-lattice decoder for all configurations of weight $d/2$ across the lattice of distances $d = 4, 6$ and $8$.

Let us next evaluate the entropic term for the unified-lattice decoder. As discussed in Sec.~\ref{ssec:motivation}, unlike the restricted-lattice decoder, the unified-lattice decoder allows us to identify correlations between the syndromes caused by Pauli-Y errors to distinguish Pauli-Y errors from Pauli-Z errors. The unified-lattice decoder fails when $d/2$ Pauli-Z errors are configured along a major row of the lattice. This occurs with probability $(1-p)^{n - d/2} (p/3)^{d/2}$. These errors may be configured along the $d$ sites of one of the $d$ major rows. We therefore obtain a logical failure rate in the low error rate limit
\begin{align}
     \overline{P}^{\textrm{uni}}_{0,n} &= (1-p)^n  \frac{d}{2}  \binom{d}{\frac{d}{2}} \left(\frac{2p }{3 (1-p)}\right)^{\frac{d}{2}} \\ &\approx  \frac{d}{2}  \binom{d}{\frac{d}{2}} \left(\frac{1}{3}\right)^{ \frac{d}{2}} p ^{\frac{d}{2}}.
\end{align}
where, as in the restricted-lattice decoder case, a factor $1/2$ is included to account for error configurations where the decoder guesses the correct solution. 
We can therefore identify the entropic term for the unified-lattice decoder
\begin{equation}
    N^{\textrm{uni}}_{\textrm{fail}} = \frac{d}{2} \binom{d}{\frac{d}{2}} (1/3)^{d/2}.
    \label{eq:N_fail_toric_unified_even}
\end{equation}
We again verify this numerically by exhaustively generating all errors of weight $d/2$ on the lattice, and decoding each error using the unified lattice for surface code distances $d = 4, 6$ and $8$, using different values of $w_A$ and $w_B$. We find that a range of weight configurations minimises the entropic term by achieving Eq.~\ref{eq:N_fail_toric_unified_even}, with the constraint that $0 < w_B < \delta$, where $\delta$ is a constant changing with code distance. We numerically find that when $w_A \approx w_B$, the entropic term is recovered. With Eq.~\ref{eq:N_fail_toric_restricted_even} and Eq.~\ref{eq:N_fail_toric_unified_even} evaluated, the ratio stated at the beginning of this subsection, Eq.~\ref{eqn:ratio} is readily checked.

\begin{figure}[t!]
    \centering
    \hspace*{-0.8cm}        
    \includegraphics[width=0.45\textwidth]{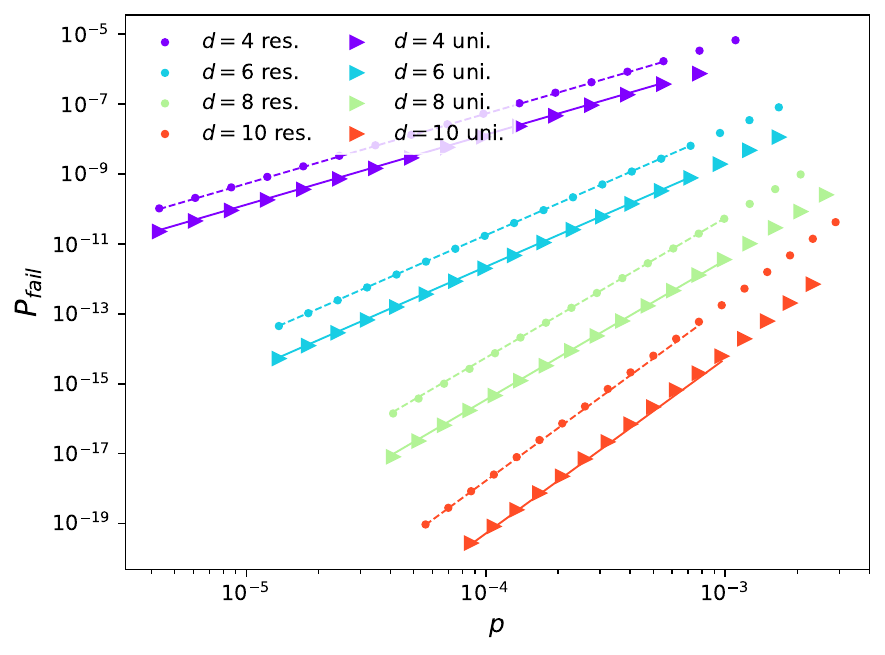}
    \caption{The logical failure rate data (in logarithmic scale) for the surface code under low physical error rates computed using the splitting method \cite{Bravyi_2013} to test the restricted and unified decoders for even surface code distances. The dashed and solid lines respectively indicate the fits obtained using Eqs.~\ref{eq:N_fail_toric_restricted_even} and \ref{eq:N_fail_toric_unified_even} for the restricted and unified decoders.} 
    \label{fig:toric_code_splitting}
\end{figure}

\subsection{\label{ssec:splitting} Low error-rate performance}

We adopt the splitting method~\cite{Bravyi_2013} for the simulation of rare events, in order to numerically investigate logical failure rates at low physical error rates $p$. These estimates are used to numerically verify the analytical expressions derived from path-counting in Sec.~\ref{ssec:path_counting_toric}. 

As discussed in Refs.~\cite{Bravyi_2013,beverland2019role}, the splitting method is used to evaluate ratios of logical failure rates $R_j = \overline{P}(p_j, n) / \overline{P}(p_{j+1}, n)$ evaluated at physical error rates $p_j$ and $p_{j+1}$ that are similar. The product of many such ratios can be used to interpolate between the high error rate regime where logical failure rates can be obtained using Monte Carlo sampling, and the path-counting regime discussed in the previous subsection where logical error rates are very small. Each term in a given ratio is evaluated by averaging over a sample of failure paths drawn from the appropriate error distribution. We note that when sampling these errors, we heuristically chose a number of steps such that statistical fluctuations are smaller than the desired logical error rate precision, and discard 50\% of the total generated Metropolis samples to ensure that the mixing
time of the Markov process was surpassed. We adopt the heuristic sequence of ratios $R_j$ with physical error rates $p_{j+1}=p_j2^{\pm1/\sqrt{w_j}}$, as proposed in Ref.~\cite{Bravyi_2013} to minimise statistical error, where $w_j=\text{max}(d/2,p_jn)$ for a code distance $d$. 

We use this splitting method to numerically interpolate between initial estimates obtained using Monte Carlo sampling at $p = 5\%$, and the analytical path counting results in the low error rate regime derived in Sec.~\ref{ssec:path_counting_toric}. Our numerical results show good agreement with our analytical estimates for the logical error rates for both the restricted and unified decoders. In Fig.~\ref{fig:toric_code_splitting} we compare logical failure rates obtained with the splitting method to our analytical expressions.

\subsection{Thresholds}\label{ssec:toric_high_ler} 

\begin{figure}[ht!]
    \centering
    \hspace*{-0.8cm}                                                     
    \includegraphics[width=0.485\textwidth]{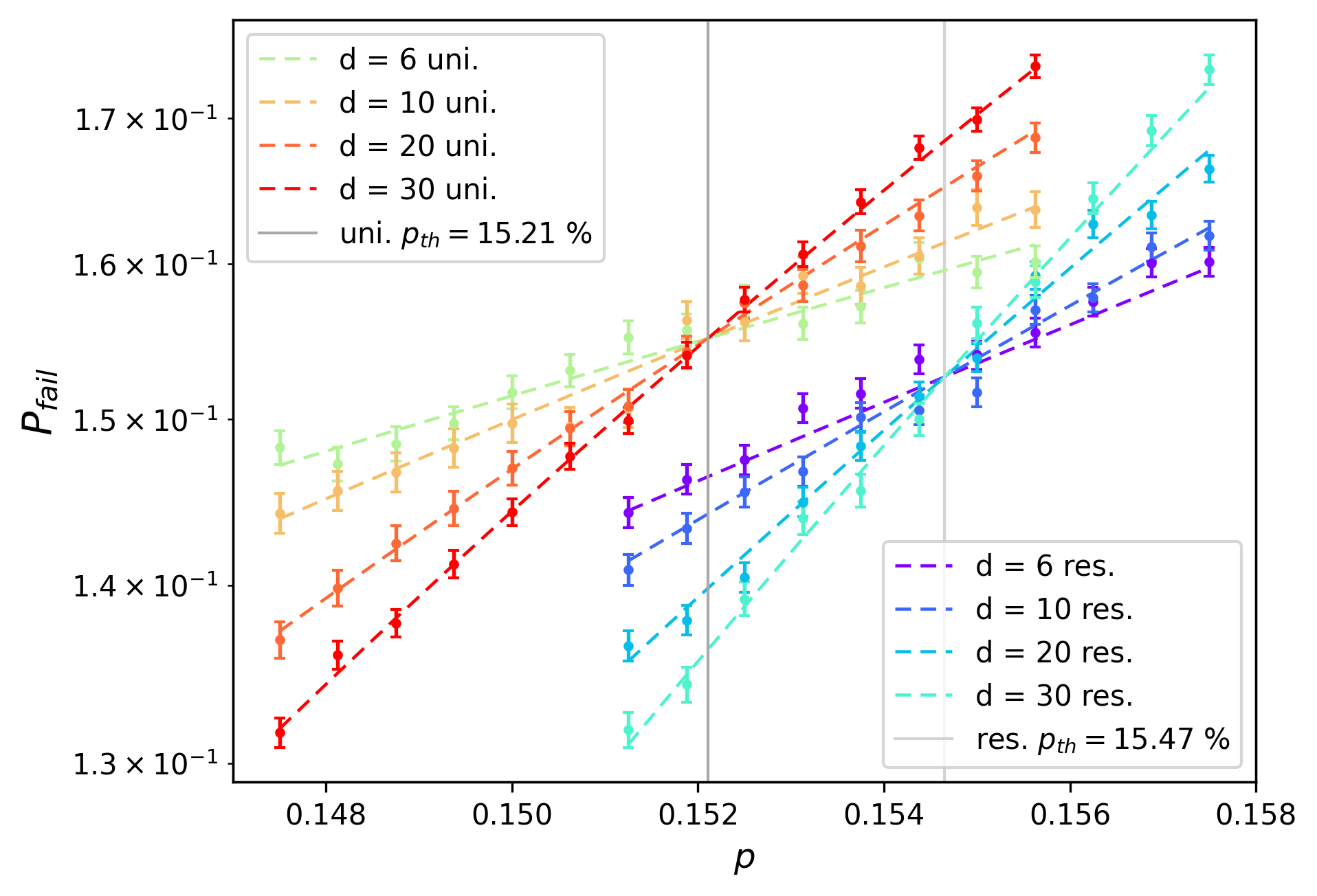} 
    \caption{The logical failure rate $P_{\textrm{fail}}$ (in logarithmic scale) as a function of physical error rate $p$ for the restricted and unified surface codes of even distance $d$, under an i.i.d depolarising noise channel. The threshold $p_{\textrm{th}}$ is indicated by the intersection. Edge weights were set to $w_A=1$ and $w_B=0.5$ in Fig.~\ref{fig:restricted_color_code}(c). Each dashed line indicates the fit to a Taylor expansion for a given system size, and the error bars show the standard deviation of the mean logical failure rate where each data point is collected using $ > 6 \times 10^4$ Monte Carlo samples.}
     \label{fig:toric_thresholds}     
\end{figure}

The threshold is a critical physical error rate $p_{\textrm{th}}$ below which a topological code is protected such that error correction succeeds with an probability exponentially suppressed as the code distance increases. Decoding  becomes ineffective above this threshold, and the logical failure rate increases with the size of the code. We fit the logical failure rate data from Monte Carlo sampling near the threshold to a second order Taylor function given by 
\begin{equation}
    f = Ax^2+Bx+C 
    \label{eq:threshold_estimate}
\end{equation}
where $x$ is the re-scaled error rate $x = (p-p_{\textrm{th}})^{1/\nu}$ \cite{wang2003confinement}. We find a fitted threshold of $p_{\textrm{th}}~\sim 15.4(7)\%$ for the restricted decoder under depolarising errors using the data shown in Fig.~\ref{fig:toric_thresholds}, where we obtain a fitted critical exponent $\nu \approx 0.611$, and 
Taylor fit constants $A = 0.471$, $B = 0.540$ and $C = 0.153$. For the unified decoder, we find that the threshold estimate varies slightly with the chosen values of $w_A$ and $w_B$. We considered multiple weights $w_B$ with fixed $w_A=1$ for edges of the matching graph in Fig.~\ref{fig:restricted_color_code}(c), and find a threshold of $p_{\textrm{th}}~\sim 15.2(1)\%$ using $w_A = 1$ for $w_B = 0.5$, as shown in Fig.~\ref{fig:toric_thresholds} where $\nu = 0.664$, $A =  0.606$, $B = 0.452$ and $C = 0.127$. We provide the data for different weight configurations in Appendix~\ref{asec:thresholds_extra_toric}, where the threshold is still comparable to that of the standard surface code, with a minor discrepancy.

\section{\label{sec:results_color_code} Color code performance under bit-flip noise}

In this section, we compare the performance of the restricted and unified decoders used to decode the color code undergoing independent and identically distributed bit-flip noise and perfect measurements, where a bit flip occurs on each qubit with probability $p$. We begin by counting the number of minimum-weight uncorrectable errors on the lattice for each decoder. We show that using the unified decoder achieves a higher success rate, providing an advantage in the low physical error-rate regime.
We verify the estimates numerically using the splitting method as was done in Sec.~\ref{ssec:splitting} \cite{Bravyi_2013}, as well as direct Monte Carlo sampling. We investigate the threshold performance of the code numerically assuming perfect measurements. 

\subsection{\label{ssec:path_counting_color} Path-counting}

We start by counting the number of minimum-weight uncorrectable error configurations on each row of the color code $n_{\textrm{fail}}^{\textrm{res}}$ of distance $D$ supporting $M=\frac{D}{2}$ square plaquettes. We define these new parameters to avoid confusion with the surface code parameters used in earlier sections. Since the lattice is square, there are $M$ such rows. Using the restricted decoder, the number of errors that cannot be corrected is given by counting configurations consisting of errors lying on such rows. The possible paths an error can take depend on the configuration of $X$ errors in an error string lying on each red square of a row, where each square can support a single-qubit $X$ error, or a two-qubit error lying either on the two vertex qubits of a horizontal edge, or on diagonal vertices of the square. Placing any of these four combinations on $\left \lfloor M/2 \right \rfloor$ locations on a row accounts for a number of strings 
\begin{equation}
n_{\textrm{fail}}^{\textrm{res}}=\sum_{k=0}^{\left \lfloor \frac{M}{2}\right \rfloor}\begin{pmatrix}
 M\\k
\end{pmatrix}\begin{pmatrix}
 M-k\\M-2k
\end{pmatrix} 4^{k}4^{M-2k},
\label{eq:color_restricted_expected_fail}
\end{equation}
where $k$ is the number of red plaquettes that support a weight-2 error, and $M-2k$ the number of weight-1 errors. Hence, the total number of logical failures on the restricted lattice corresponds to the number of equivalent configurations supported on all $M$ rows, namely
\begin{equation} 
\begin{split}
N_{\textrm{fail}}^{\textrm{res}} & = \frac{1}{2}n_{\textrm{fail}}^{\textrm{res}}M \\
 & = \frac{M}{2}\sum_{k=0}^{\left \lfloor \frac{M}{2}\right \rfloor}\begin{pmatrix}
 M\\k
\end{pmatrix}\begin{pmatrix}
 M-k\\k
\end{pmatrix} 4^{M-k}
\end{split} 
\label{eq:N_fail_color_restricted} 
\end{equation}
where the prefactor of one half is due to the classical decoder misidentifying the right correction with a $50\%$ probability.

We consider the MWPM problem on the unified lattice shown in Fig.~\ref{fig:restricted_color_code}(c). Unlike decoding on the restricted lattice, a diagonal error gives rise to a syndrome mapped onto all sub-lattices of the unified lattice, as depicted in Fig.~\ref{fig:example_of_unified_decoder_success}. On each row supporting $M$ square plaquettes, the number of weight-$M$ error configurations that cannot be corrected is given by
\begin{equation}
n_{\textrm{fail}}^{\textrm{uni}}=\sum_{k=0}^{\left \lfloor \frac{M}{2}\right \rfloor}\begin{pmatrix}
 M\\k
\end{pmatrix}\begin{pmatrix}
 M-k\\M-2k
\end{pmatrix} 2^{k}4^{M-2k},
\end{equation}
where $k$ is the number of weight-2 errors that can reside on each square plaquette. As for the restricted lattice, the total number of logical failures when decoding on the unified lattice is 
\begin{equation} 
\begin{split}
N_{\textrm{fail}}^{\textrm{uni}}& = \frac{1}{2}n_{\textrm{fail}}^{\textrm{uni}}M \\
 & = \frac{M}{2}\sum_{k=0}^{\left \lfloor \frac{M}{2}\right \rfloor}\begin{pmatrix}
 M\\k
\end{pmatrix}\begin{pmatrix}
 M-k\\k
\end{pmatrix} 4^{M-\frac{3}{2}k}.
\end{split}
\label{eq:N_fail_color_unified}
\end{equation} 
We numerically verified these estimate by exhaustively generating all error strings of weight $M$ on each $M$-square row of the color code, mapping the error configuration respectively to a restricted and unified lattice, and running the MWPM decoder.

\subsection{\label{ssec:low_error_rate_color} Low error-rate performance}

\begin{figure}[t]
    \centering   
    \hspace*{-0.8cm}                                                             \includegraphics[width=0.485\textwidth]{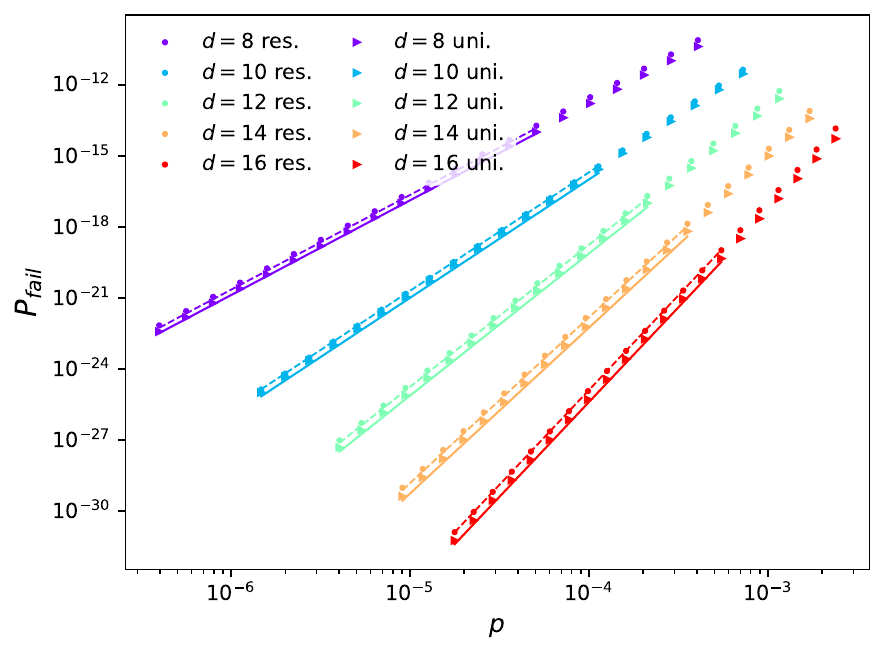}
    \caption{Logical failure rates at low physical error rates (in logarithmic scale) computed using the splitting method \cite{Bravyi_2013} with the restricted and unified decoders on the color code under i.i.d bit-flip noise. For the latter, edge weights were set to $w_A=1.1, w_B=1$. The dashed and solid lines respectively indicate the fits using Eq.~\ref{eq:N_fail_color_restricted} and ~\ref{eq:N_fail_color_unified}.} 
    \label{fig:color_code_splitting}
\end{figure} 

We implement the splitting method under the assumption of errors drawn from an i.i.d bit-flip noise on the color code, which can be written as 
\begin{equation}
    \pi_j(E) = p_j^{|E|}(1-p_j)^{n-|E|}
    \label{eq:error_pdf_color_code}
\end{equation}
where $n$ is the number of physical qubits in the color code, $|E|$ the weight of an error operator $E$. Once again, we adopt the sequence $p_{j+1}=p_j2^{\pm1/\sqrt{w_j}}$ to split ratios~\cite{Bravyi_2013} to minimise statistical error, where $w_j=\text{max}(d/2,p_jn)$ for a code distance $d$. We numerically interpolate between initial estimates obtained using Monte Carlo sampling at physical error rate $p = 5\%$, and the analytical path counting results in the low error rate regime shown in Sec.~\ref{ssec:path_counting_color}. The results are shown in Fig.~\ref{fig:color_code_splitting} where a MWPM decoder was executed on the color code restricted and unified lattices respectively, under bit-flip noise and perfect measurements. For the distances considered, the analytic estimates in Eqs.~\ref{eq:N_fail_color_restricted} and \ref{eq:N_fail_color_unified} accurately predict the dominant behaviour of the error chains in the low error rate regime, as indicated by the solid line fits. We note that in order to correct for all expected minimum-weight configurations for bit-flip errors on the color code there is a constraint on the edge weights, namely that $w_A > w_B$, which does not exist in the case of depolarising errors on the even-distance surface code.

\subsection{\label{ssec:thresh_color} Thresholds}

\begin{figure}[ht!]
    \centering
    \hspace{-5mm}\includegraphics[width=0.48\textwidth]{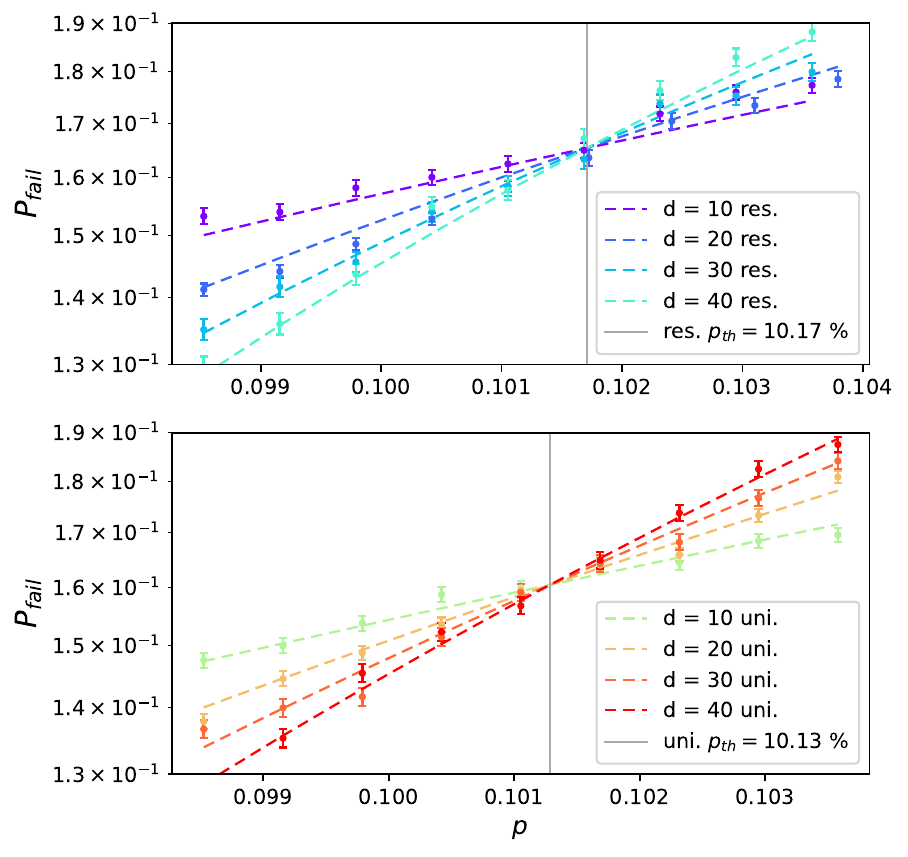}
    \caption{The logical failure rate $P_{\textrm{fail}}$ (in logarithmic scale) as a function of physical error rate $p$ for codes of different distance $d$, under an i.i.d bit-flip noise model. The threshold indicated by the intersection is found to be $p_{\textrm{th}} \sim 10.2\%$ for the restricted decoder (top). The unified decoder threshold (bottom) is $p_{\textrm{th}} \sim 10.1(3)\%$, using edge weights $w_A = 2.1$ and $w_B = 1$ with the matching graph in Fig.~\ref{fig:restricted_color_code}(c). Each dashed line indicates this fitting for a given system size, and the error bars show the standard deviation of the mean logical failure rate where each data point is collected using $ > 6 \times 10^4$ Monte Carlo samples.}
    \label{fig:color_code_thresholds}
\end{figure}

The restricted decoder recovers the threshold $p_{\textrm{th}}~\sim 10.2\%$ under independent and identically distributed bit-flip noise, see Fig.~\ref{fig:color_code_thresholds}, where we obtain a fitted critical exponent $\nu = 0.628$, and 
Taylor fit constants $A =-0.0419$, $B = 1.136$ and $C = 0.165$ for Eq.~\ref{eq:threshold_estimate}. This is consistent with that of the surface code on a square lattice \cite{wang2003confinement, delfosse2014decoding}. In the case of the unified decoder, we considered multiple weights $w_A$ with $w_B=1$ for edges of the matching graph in Fig.~\ref{fig:restricted_color_code}(c), and find a peak in threshold at $p_{\textrm{th}} \sim 10.1(3)\%$ around $w_A = 2.1$ (Fig.~\ref{fig:color_code_thresholds}). Surprisingly, the unified decoder demonstrates a marginally lower threshold than the restricted decoder. This pattern is consistent with the threshold comparison applied to the surface code in Sec.~\ref{ssec:toric_high_ler}. We hypothesize that in the near-threshold regime, the presence of additional defects on the unified lattice contributes to an increased number of incorrect minimum-weight paths that the decoder may choose, leading to this reduction.

\section{\label{sec:discussion} Discussion}

In conclusion, we demonstrated a method to reduce the logical error rate of the surface code under depolarising errors. This was done by first outlining a mapping from the surface to the color code. We then showed that this noise model on the surface code can be mapped onto a set of two-qubit bit-flip errors on the color code. Subsequently, we constructed a unified lattice in the form of a punctured torus, for use under a minimum-weight perfect matching decoder to correct high-weight errors. To evaluate the
performance of our decoder we analytically and numerically investigated the logical failure rates at low physical error rates and verified their agreement. For even code distances, our unified decoder achieves a least-weight correction for all weight $d/2$ depolarising errors on the surface code. We also demonstrated an exponential suppression in the logical error rate for the color code under independent and identically distributed bit-flip noise, extending the work in Ref.~\cite{sahay2022decoder} to the square-octagonal color code with boundaries.

It will be interesting to determine how the performance of our decoder compares to other practical decoders that are designed to account for depolarising noise, both in terms of error-correction performance and runtime. For instance, both iterative~\cite{delfosse2014decoding, fowler2013optimal} and belief-matching decoders~\cite{higgott2023improved} have been proposed to augment matching decoders with the ability to take advantage of correlations between the syndromes measured by the star and plaquette operators. To complete such a comparison, it will first be necessary to determine the performance of other decoding algorithms in the limit of low error rate. Of course, the matching decoder we have presented here is compatible with other decoding techniques such as belief-matching and two-stage decoding \cite{criger2018multi, roffe2023bias}, at the cost of added computational complexity, and it may be interesting to design versatile decoders for other codes by adapting these techniques. However, it is important to note that our decoder only requires a single invocation of an MWPM decoder to address Pauli correlations. As a result, it benefits from the expected linear time complexity of MWPM \cite{higgott2023sparse}, with only a constant factor increase of 4 in the size of the matching graph. One could also readily combine our decoder with fast decoders such as union-find~\cite{Delfosse2021almostlineartime, chan2023strictly, griffiths2023unionfind} or weighted union-find~\cite{Huang2020} to improve its runtime. We note that we can also generalise our versatile matching decoder to deal with non-trivial syndrome correlations that occur in fault-tolerant error correction~\cite{wang2003confinement, Raussendorf2005long, Fowler12surface}, and may be adapted for use in other variations of the surface code with boundaries such as Floquet codes~\cite{Haah2022boundarieshoneycomb, kesselring2022anyon}. One might also explore generalising our decoding method to surface codes with alternative topologies, or perhaps using other color code symmetries. Both of these directions are discussed in Appendix~F of Ref.~\cite{sahay2022decoder}.

Broadly speaking, we have shown that we can obtain a matching decoder for the surface code capable of correcting depolarising noise by mapping its syndrome onto an enlarged state space, namely, the syndrome of the color code. It will be interesting to explore this method further to determine the extent to which it generalises. For instance, it is compelling to investigate whether we obtain near-optimal performance using our mapped decoder for different lattice geometries~\cite{beverland2019role}. 
Another important case to consider is the circuit noise model, where decoding is conducted on a $(2+1)$-dimensional syndrome history. Trivially, we can adapt the matching graph obtained from the unified lattice for this setting by extending it along the time direction in the standard way~\cite{dennis2002topological, brown2020parallelized, brown2022conservation}. Nevertheless, it will be very interesting if the novel decoding methods we have proposed can be generalised to offer an advantage in correcting the types of correlated errors the circuit noise model can introduce. Perhaps, for instance, the edge weights on the unified lattice can be chosen appropriately to account for correlated errors such as hook errors. Perhaps there is even some other alternative lattice that can be found to offer an advantage in dealing with this more complicated noise model.
More generally, it is exciting to consider other decoding problems on extended Hilbert spaces. This may lead to practical, high-performance decoders for more sophisticated codes, as well as more exotic correlated noise models. We leave this exploration to future work.

\begin{acknowledgments} We thank N. Delfosse for helpful conversations. A.B. and K.S. are grateful to the organizers of the IBM Quantum Error Correction Summer School where parts of this collaboration took place. 
A.B. acknowledges funding from the EPSRC Centre for Doctoral Training in Delivering Quantum Technologies at UCL, Grant No. EP/S021582/. B.J.B. received support from the European Union’s Horizon 2020 research and innovation programme under the Marie Skłodowska-Curie grant agreement No. 897158. B.J.B. changed affiliation to IBM Quantum during the preparation of the present manuscript.
\end{acknowledgments}

\section*{Data Availability}
The data that support the findings of this study are available from the corresponding author upon reasonable request. The source code and data used in the figures in this manuscript can be accessed at \url{https://github.com/abenhemou/unimatch}.

\bibliographystyle{quantum}
\bibliography{references.bib}

\onecolumn
\newpage
\appendix

\section{Pseudocode for unified decoding on the surface code}

Here, we outline the Unified decoder presented in the main text. The procedure requires a set of syndrome measurements of the color code stabilizers $\sigma_{\textrm{c.c}}$, a subroutine which restricts $\sigma_{\textrm{c.c}}$ to syndromes on restricted lattices, and a subroutine which builds the unified lattice as prescribed in Fig.~\ref{fig:restricted_color_code}. A minimum-weight perfect matching decoder \texttt{mwpmdecoder} is then used to decode the syndrome and provide a correction $\bar{C}$. The correction operator returns the code to the codespace, and its success is given by the condition that the decoding algorithm has determined that the parity of the edges in the matching solution crossing a chosen logical representative $\bar{Z}$ is equal to the parity of Pauli errors which occurred on qubits $q \in \bar{Z}$, as mentioned in Sec.~\ref{ssec:decoding_on_unified_lattice}.

\begin{algorithm}
    \caption{Unified Decoder}
    \label{alg:ufdecoder}

     \textbf{Require:} \newline Restricted lattices $\mathcal{R}_{\mathbf{u}} = \{ \{S_f \}_{\mathbf{col}(f) \neq \mathbf{u} } \}$ on a manifold $G_{\mathcal{R}_{\mathbf{u}}}$  \newline Unified lattice constructor $\mathcal{U}(.)$ \newline A MWPM decoder \texttt{mwpmdecoder} 
     
    \textbf{Input:} A color code syndrome $\sigma_{\textrm{c.c.}}$ on the color code manifold $G_{\textrm{c.c}}$ 

    \textbf{Output:} A correction operator $\bar{C}$ 

    \tcc{Map syndromes to restricted lattices}
    
    \ForEach{$\textbf{\textrm{u}} \in \{\textbf{\textrm{r}},
    \textbf{\textrm{g}},
   \textbf{\textrm{b}}\}$}
    {
    
    \ForEach{$\sigma_i \in \sigma_{\textrm{c.c.}}$}
        {
            \If{$ color(\sigma_i) = \textbf{\textrm{u}}$}
            {
                % $\mathcal{R}_{\mathbf{u}}(\sigma_{i})$ on $G_{\mathcal{R}_{\mathbf{u}}}$ \gets $\sigma_i$ on $G_{\textrn{c.c}}$ 

                $\mathcal{R}_{\mathbf{u}}(\sigma_i) \text{ on } G_{\mathcal{R}_{\mathbf{u}}} \gets \sigma_i \text{ on } G_{\textrm{c.c.}}$
                
            }
        }
    }

    \tcc{Construct unified lattice, and map syndrome on the unified lattice}

    % $( G_{\mathcal{U}}, \sigma_{i,\mathcal{U}} )$ \gets $ \ensuremath{\mathcal{U}}(G_{\mathcal{R}_{\mathbf{r}}}, 
    % G_{\mathcal{R}_{\mathbf{g}}}, 
    % G_{\mathcal{R}_{\mathbf{b}}},
    % \mathcal{R}_{\mathbf{r}}(\sigma_i), \mathcal{R}_{\mathbf{g}}(\sigma_i), \mathcal{R}_{\mathbf{b}}(\sigma_i))$

    $(G_{\mathcal{U}}, \sigma_{i,\mathcal{U}}) \gets \mathcal{U}\left(
        G_{\mathcal{R}_{\mathbf{r}}}, 
        G_{\mathcal{R}_{\mathbf{g}}}, 
        G_{\mathcal{R}_{\mathbf{b}}},
        \mathcal{R}_{\mathbf{r}}(\sigma_i), 
        \mathcal{R}_{\mathbf{g}}(\sigma_i), 
        \mathcal{R}_{\mathbf{b}}(\sigma_i)
    \right)$

    \tcc{Decode}
     $ \bar{C} = \texttt{mwpmdecoder}( G_{\mathcal{U}}, \sigma_{i,\mathcal{U}})$
    
    \tcc{Apply Pauli-$Z$ corrections to qubits along the shortest paths determined by the decoder}
    
    Apply $\prod_{q\in\bar{C}} Z_q$

\end{algorithm}

\newpage
\section{\label{asec:thresholds_extra_toric} Thresholds for the even-distance surface code using varied weights}

\begin{figure}[ht!]
    \centering
    \subfigure[$w_B=0.7$]{\includegraphics[width=0.34\textwidth]{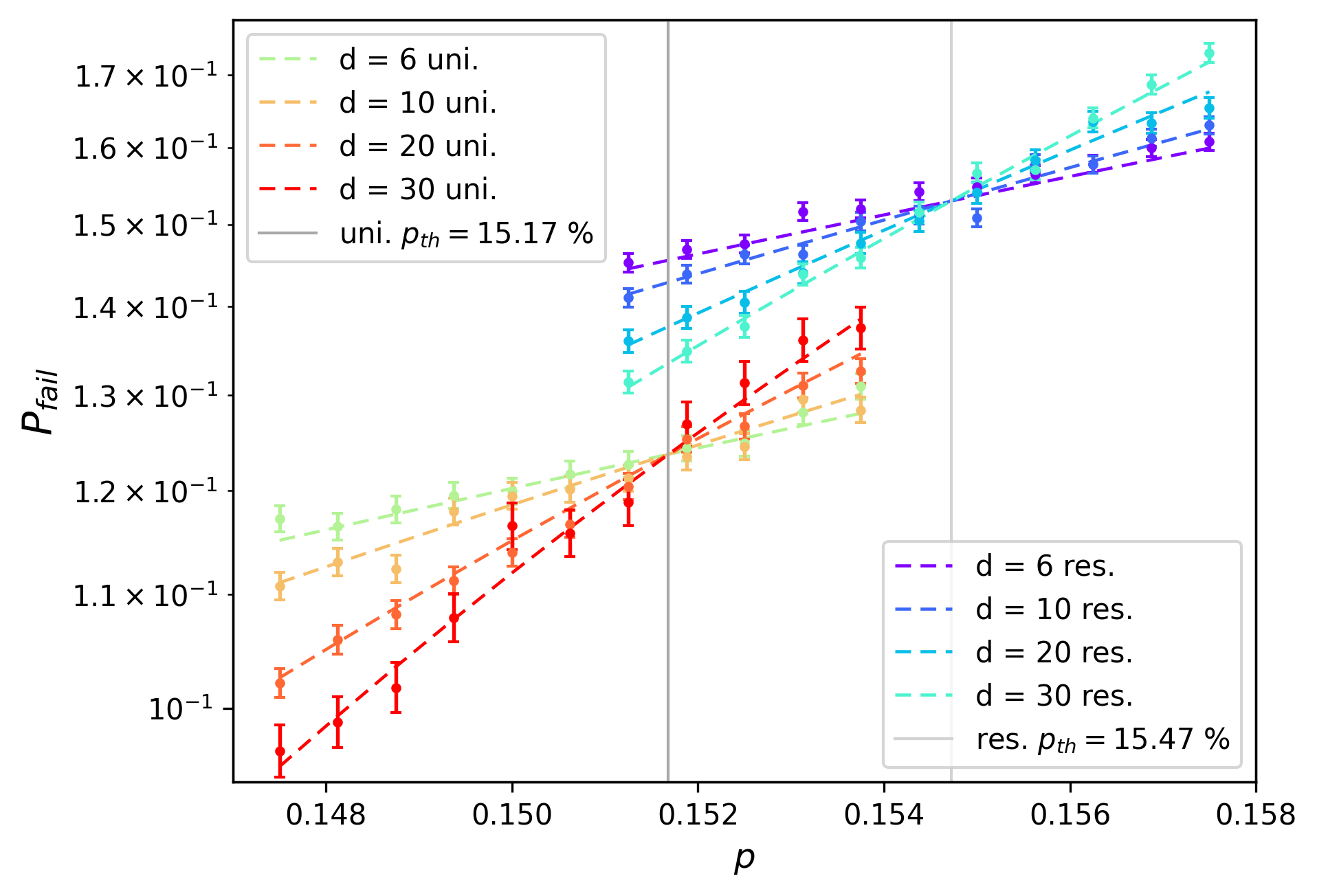}} 
    \subfigure[$w_B=1.0$]{\includegraphics[width=0.34\textwidth]{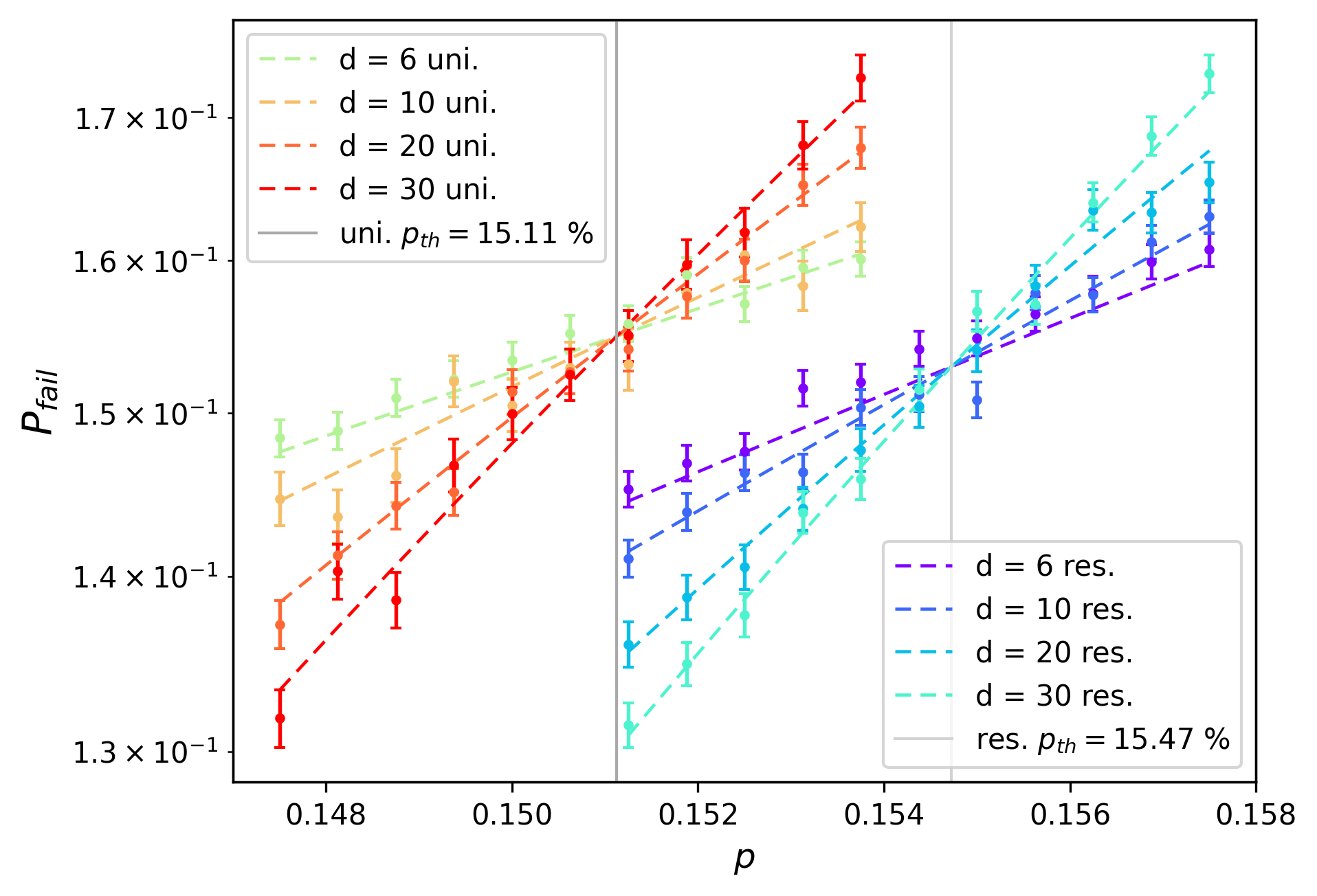}} 
    \subfigure[$w_B=1.1$]{\includegraphics[width=0.34\textwidth]{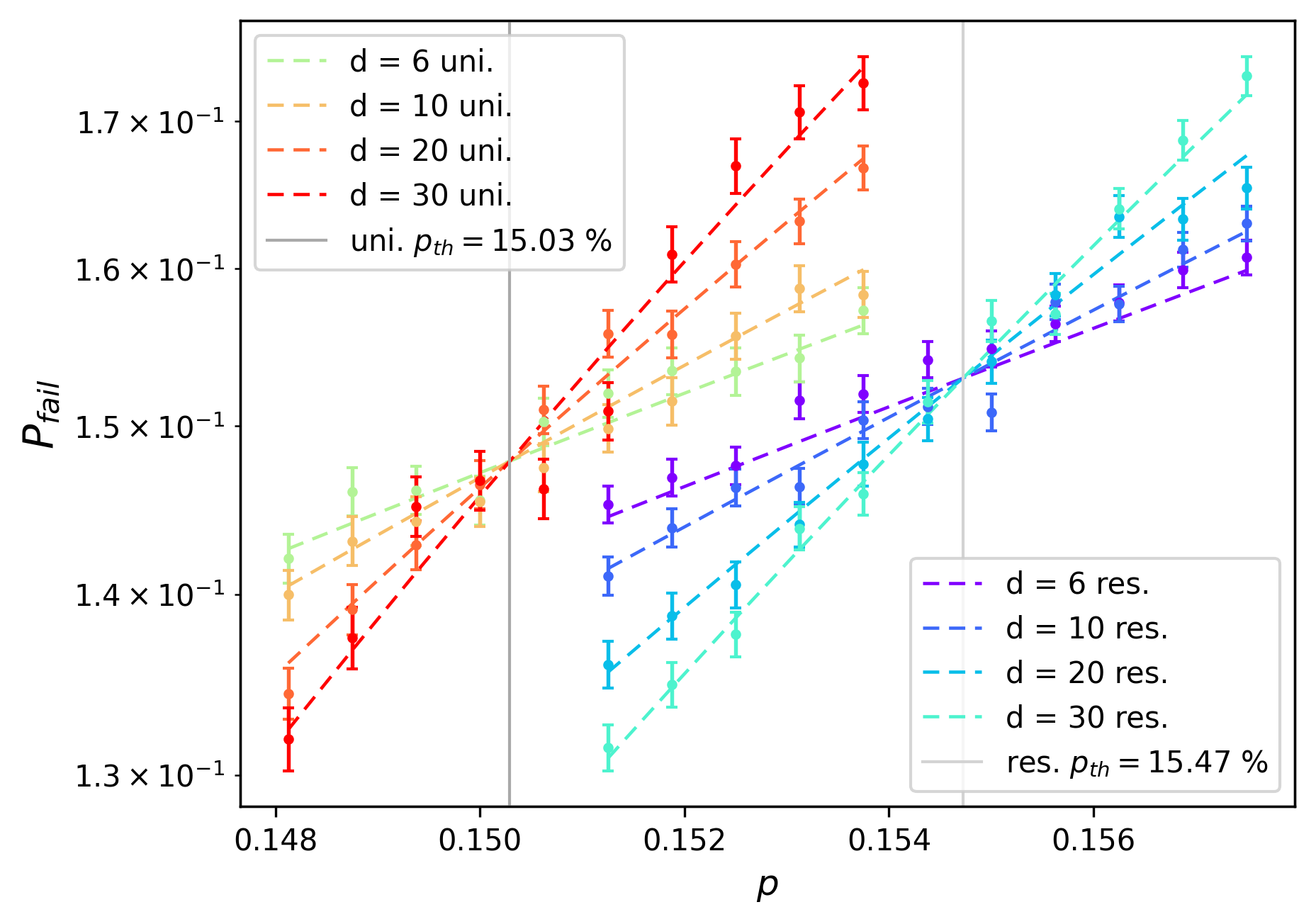}}
    \subfigure[$w_B=2$]{\includegraphics[width=0.34\textwidth]{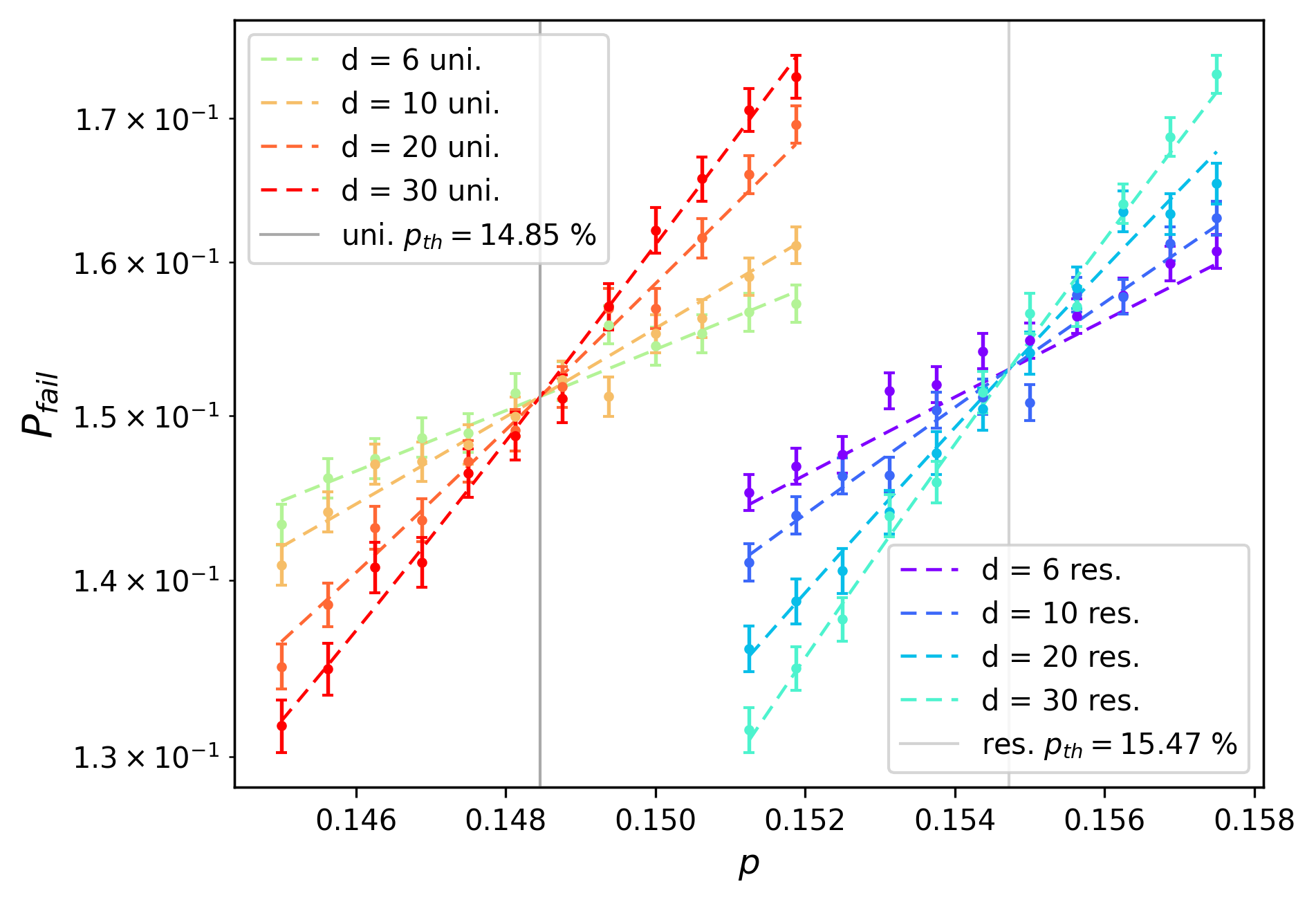}} 
    \label{fig:toric_unified_threshold_varied_weights} 
    \caption{The logical failure rate $P_{\textrm{fail}}$ (in logarithmic scale) as a function of physical error rate $p$ for surface codes of different even distance $d$, under i.i.d depolarising noise. The restricted decoder threshold is shown against the unified decoder configured with weights $w_A=1$, and (a) $w_B=0.7$, (b) $w_B=1$ and (c) $w_B=1.1$, (d) $w_B=2$ using the matching graph in Fig.~\ref{fig:restricted_color_code}(c). Each dashed line indicates the fit to a Taylor expansion for each system size, and the error bars show the standard deviation of the mean logical failure rate where each data point is collected using $ > 6 \times 10^4$ Monte Carlo samples.}
\end{figure}

\end{document}